%% file: delayedsubst.tex
\documentclass[submission,copyright,creativecommons]{eptcs}
\providecommand{\event}{F-IDE 2021} 
\usepackage{breakurl}             
\usepackage{underscore}           

\usepackage{graphicx}
\usepackage{wrapfig}

\usepackage{amsmath}
\usepackage{mathtools}
\RequirePackage{stmaryrd}
\usepackage{fontawesome}

\usepackage{cite}

\usepackage{hyperref}
\hypersetup{
    colorlinks,
    linkcolor={red!50!black},
    citecolor={blue!50!black},
    urlcolor={blue!80!black}
}

\usepackage{xcolor}

\usepackage{paralist}

\usepackage{math}
\usepackage[prefixflatinterpret,bracketmodalinterpret,fixformat,silentconst,sidenotecalculus,longseqcontext]{logic}
\usepackage[prefixflatinterpret,bracketmodalinterpret,fixformat,silentconst,differentialdL,simplenames]{dL}
\usepackage[irlabel]{fcps}

\usepackage{subcaption}

\usepackage{booktabs}
\usepackage{tabularx}
\usepackage{multirow}

\usepackage{tikz}
\usepackage{pgfmath}
\usetikzlibrary{decorations.text,arrows.meta,calc,shadows.blur,shadings,positioning,patterns,fit}

\usepackage[normalem]{ulem}

\usepackage[T1]{fontenc}
\usepackage[scaled=0.85]{beramono}

\usepackage{prettyref}
\newcommand{\rref}[2][]{\prettyref{#2}}
\newrefformat{sec}{Section\,\ref{#1}}
\newrefformat{def}{Def.\,\ref{#1}}
\newrefformat{thm}{Theorem\,\ref{#1}}
\newrefformat{prop}{Proposition\,\ref{#1}}
\newrefformat{lem}{Lemma\,\ref{#1}}
\newrefformat{cor}{Corollary\,\ref{#1}}
\newrefformat{ex}{Example\,\ref{#1}}
\newrefformat{tab}{Table\,\ref{#1}}
\newrefformat{fig}{Fig.\,\ref{#1}}
\newrefformat{app}{Appendix\,\ref{#1}}
\newrefformat{lstl}{Line\,\ref{#1}}

\definecolor{vgreen}{rgb}{.1,.5,0}
\definecolor{vred}{rgb}{.7,0,0}
\definecolor{vblue}{rgb}{.1,.15,.62}

\usepackage{listings}
\usepackage{keylistings}

\lstdefinelanguage{KeYmaeraX}{%
  keywords={if,then,else,Real,Bool,HP,Definitions,ProgramVariables,Problem,End,Tactic,Lemma,ArchiveEntry,Axiom,@invariant},  sensitive=true,
  morecomment=[s]{/*}{*/},
  deletestring=[d]',
  morestring=[b]",
  showstringspaces=false,
  commentstyle=\color{vgreen},
  mathescape,
  escapeinside={/*@}{@*/},
  basicstyle=\small\ttfamily\upshape,
  stringstyle=\color{vgreen}}[keywords]

\lstdefinelanguage{Bellerophon}{%
  language={},
  keywords={'R,'L,'_},%
  otherkeywords={;,<,|},
  sensitive=true,
  morecomment=[l]{//},
  morecomment=[s]{/*}{*/},
  morestring=[b]",
  deletestring=[d]',
  morestring=[d]`,
  showstringspaces=false,
  commentstyle=\color{vgreen}}[keywords]

\newcommand*\circled[1]{\tikz[baseline=(char.base)]{
            \node[shape=circle,draw,inner sep=2pt] (char) {#1};}}
\newcommand*\bcircled[1]{\tikz[baseline=(char.base)]{
            \node[shape=circle,draw,inner sep=2pt,fill=black,text=white] (char) {\textbf{#1}};}}

\title{Implicit and Explicit Proof Management in \KeYmaeraX
\thanks{This material is based upon work supported by the Air Force Office of Scientific Research under grant number FA9550-16-1-0288 and FA8750-18-C-0092.  Any opinions, finding, and conclusion or recommendations expressed in this material are those of the author(s) and do not necessarily reflect the views of the United States Air Force.}
}
\author{Stefan Mitsch
\institute{Computer Science Department\\
Carnegie Mellon University\\
Pittsburgh, PA, USA}
\email{smitsch@cs.cmu.edu}
}
\def\titlerunning{Proof Management in \KeYmaeraX}
\def\authorrunning{S. Mitsch}
\begin{document}
\maketitle

\begin{abstract}
Hybrid systems theorem proving provides strong correctness guarantees about the interacting discrete and continuous dynamics of cyber-physical systems.
The trustworthiness of proofs rests on the soundness of the proof calculus and its correct implementation in a theorem prover.
Correctness is easier to achieve with a soundness-critical core that is stripped to the bare minimum, but, as a consequence, proof convenience has to be regained outside the soundness-critical core with proof management techniques.
We present modeling and proof management techniques that are built on top of the soundness-critical core of \KeYmaeraX to enable expanding definitions, parametric proofs, lemmas, and other useful proof techniques in hybrid systems proofs.
Our techniques steer the uniform substitution implementation of the differential dynamic logic proof calculus in \KeYmaeraX to allow users choose when and how in a proof abstract formulas, terms, or programs become expanded to their concrete definitions, and when and how lemmas and sub-proofs are combined to a full proof.
The same techniques are exploited in implicit sub-proofs (without making such sub-proofs explicit to the user) to provide proof features, such as temporarily hiding formulas, which are notoriously difficult to get right when implemented in the prover core, but become trustworthy as proof management techniques outside the core.
We illustrate our approach with several useful proof techniques and discuss their presentation on the \KeYmaeraX user interface.
\end{abstract}


\irlabel{DUS|$\circlearrowright$US}%


\section{Introduction}
\label{sec:introduction}

Hybrid systems theorem proving provides strong correctness guarantees about the interacting discrete and continuous dynamics of cyber-physical systems.
Theorem proving is most valuable early in the design of a system, since it is not merely a technique to prove the correctness of an already correct system, but also shines when analyzing a system in all its subtleties to discover unknown or only partially known properties of the system.
The trustworthiness of proofs and analysis results, however, rests on the soundness of the proof calculus and its correct implementation in a theorem prover.
Typical theorem prover implementations often opt for directly representing the rules of a proof calculus in the theorem prover, for instance, with axiom schemata in~\cite{KeY2005,DBLP:conf/cade/PlatzerQ08}, or with trusted implementations of rules~(e.g., KeYmaeraD~\cite{DBLP:conf/icfem/RenshawLP11}) or decision procedures~(e.g., invariant computation~\cite{DBLP:conf/emsoft/LiuZZ11} in the HHL prover~\cite{DBLP:conf/icfem/WangZZ15}).
The downside of such an approach is not only that implementations of rules and their side conditions become soundness-critical, but also that additional features often result in increasing the size of the soundness-critical code base of the theorem prover.
Correctness is easier to achieve with an LCF-style approach that strips the soundness-critical core to the bare minimum, but, as a consequence, proof convenience has to be regained outside the soundness-critical core with proof management techniques.
The \KeYmaeraX~\cite{DBLP:conf/cade/FultonMQVP15} theorem prover for hybrid systems takes an LCF-style approach; previous techniques expanded the capabilities of \KeYmaeraX primarily by providing tactics~\cite{DBLP:conf/itp/FultonMBP17}, e.g., for certifying solutions of differential equations~\cite{DBLP:journals/jar/Platzer17}, for certifying safety and liveness properties of differential equations~\cite{DBLP:journals/jacm/PlatzerT20,DBLP:journals/fmsd/SogokonMTCP}, for stability proofs~\cite{DBLP:conf/tacas/TanP21}, for code synthesis~\cite{DBLP:conf/pldi/BohrerTMMP18}, for component-based modeling and verification~\cite{DBLP:journals/sttt/MullerMRSP18}, and for monitor synthesis~\cite{DBLP:journals/fmsd/MitschP16}.

In this paper, we present modeling and proof management techniques that are built on top of the soundness-critical core of \KeYmaeraX to enable structuring and modularizing models with definitions and modularizing proofs with lemmas.
These modeling and proof management techniques were developed primarily with interactive proofs in mind, but may also be beneficial for automation (e.g., hierarchical definitions may serve as proof hints).
Useful proof techniques for explicit proof management include expanding definitions of the model during a proof, parametric proofs to make progress in proofs despite unknown system properties (e.g., loop invariants), and creating and applying lemmas.
Our techniques steer the uniform substitution implementation of the differential dynamic logic proof calculus in \KeYmaeraX to allow users choose when in a proof and how abstract formulas, terms, or programs become expanded to their concrete definitions, and when and how lemmas and sub-proofs are combined to a full proof.
The same techniques are exploited in implicit sub-proofs (without making such sub-proofs explicit to the user) to hide technicalities of the prover implementation whose details are irrelevant to the user, or to provide proof features, such as temporarily hiding formulas, which are notoriously difficult to get right when implemented in the prover core, but become trustworthy as proof management techniques outside the core.
On the user interface, we attempt to make such proof features available as part of the usual user interactions: for example, when a tactic asks for input (e.g., a loop invariant), users start a parametric proof simply by using uninterpreted function and predicate symbols as tactic inputs, which then appear like elements of the input model whose concrete interpretations can be defined and expanded at a later point in the proof.
That way, users can focus on exploring and understanding a system by way of formal proof to provide insight to the theorem prover when it becomes available during the proof.

The remainder of this paper is structured as follows:
\rref{sec:preliminaries} introduces differential dynamic logic and the relevant core and user interface features of \KeYmaeraX.
\rref{sec:proofmanagementexample} gives an example proof that combines and illustrates several of the desired proof management techniques, \rref{sec:lemmas} and \rref{sec:prooftechniques} discuss the underlying lemma application and proof techniques and their appearance on the user interface, and \rref{sec:conclusion} concludes the paper with a discussion of related and future work.

\section{Preliminaries}
\label{sec:preliminaries}

\paragraph{Differential Dynamic Logic by Example}

Differential dynamic logic \dL~\cite{DBLP:journals/jar/Platzer17,Platzer18} is a specification language and verification calculus for hybrid systems written as hybrid programs.
The syntax of \emph{hybrid programs} (HP) is described by the following grammar where $\asprg,\bsprg$ are hybrid programs, $x$ is a variable and $\astrm,\genDE{x}$ are terms, $\ivr$ is a logical formula:
\begin{equation*}
  \asprg,\bsprg ~\bebecomes~
  \pupdate{\pumod{x}{\astrm}}
  \alternative
  \ptest{\ivr}
  \alternative
  \pevolvein{\D{x}=\genDE{x}}{\ivr}
  \alternative
  \pchoice{\asprg}{\bsprg}
  \alternative
  \asprg;\bsprg
  \alternative
  \prepeat{\asprg}
\end{equation*}

Assignments \(\pupdate{\pumod{x}{\astrm}}\) and tests \(\ptest{\ivr}\) (to abort execution and discard the run if $\ivr$ is not true) are as usual.
Differential equations \(\pevolvein{\D{x}=\genDE{x}}{\ivr}\) are followed along a solution of \(\D{x}=\genDE{x}\) for any duration as long as the evolution domain constraint $\ivr$ is true at every moment along the solution.
Nondeterministic choice \(\pchoice{\asprg}{\bsprg}\) runs either $\asprg$ or $\bsprg$, sequential composition \(\asprg;\bsprg\) first runs $\asprg$ and then $\bsprg$ on the resulting states of $\asprg$, and nondeterministic repetition \(\prepeat{\asprg}\) runs $\asprg$ any natural number of times.
For example, the hybrid program below
\[
\prepeat{\bigl(\underbrace{\text{if}~(x<1) \{\humod{y}{-x}\}~\text{else}~\{\humod{y}{*};\ptest{y>2}\}}_\textit{ctrl};~\underbrace{\pevolve{\D{x}=-xy}}_\textit{ode}\bigr)}
\]
repeats program $\textit{ctrl}$ followed by differential equation $\textit{ode}$ arbitrarily often; program $\textit{ctrl}$ is a choice between setting $y$ to the value of $-x$ when $x<1$ or else picking any $y>2$.
The combined effect of $\textit{ctrl}$ and $\textit{ode}$ is an exponential increase/decay of $x$ with a rate depending on the choice of $y$.
When programs become more complicated, it is useful to literally modularize hybrid programs into $\textit{ctrl}$, $\textit{ode}$ etc. using program symbols and use definitions as a structuring mechanism for models.

The formulas of \dL describe properties of hybrid programs, summarized by the following grammar
  where $\asfml,\bsfml$ are formulas, $\astrm,\bstrm$ are terms, $x$ is a variable and $\asprg$ is a hybrid program:
\begin{equation*}
  \asfml,\bsfml ~\bebecomes~
  \astrm\geq\bstrm \alternative
  \lnot \asfml \alternative
  \asfml \land \bsfml \alternative
  \asfml \lor \bsfml \alternative
  \asfml \limply \bsfml \alternative
  \asfml \lbisubjunct \bsfml \alternative
  \lforall{x}{\asfml} \alternative
  \lexists{x}{\asfml} \alternative
  \dbox{\asprg}{\asfml}
  \alternative \ddiamond{\asprg}{\asfml}
\end{equation*}
The operators of first-order real arithmetic are as usual with quantifiers ranging over the reals.
Formula \(\dbox{\asprg}{\asfml}\) is true in a state iff formula $\asfml$ is true after all ways of running hybrid program $\asprg$, which is useful for expressing safety properties.
Dually, liveness properties are expressed with \(\ddiamond{\asprg}{\asfml}\), which is true in a state iff $\asfml$ is true after at least one run of $\asprg$.

\paragraph{Proofs in the \KeYmaeraX Core}

\begin{figure}[htb]
\begin{subfigure}[b]{\textwidth}
\begin{sequentdeduction}[array]
\linfer
{\lsequent{\Gamma_1}{\Delta_1~(\text{subgoal}_1)}
!\ldots
!\lsequent{\Gamma_n}{\Delta_n~(\text{subgoal}_n)}
}
{\lsequent{\Gamma}{\Delta~(\text{conclusion})}}
\end{sequentdeduction}
\caption{A \texttt{Provable} representing proof state in the \KeYmaeraX core}
\label{fig:proofstate-provable}
\end{subfigure}
\begin{subfigure}[b]{\textwidth}
\begin{tikzpicture}
\node (ui) {
\includegraphics[width=.8\textwidth]{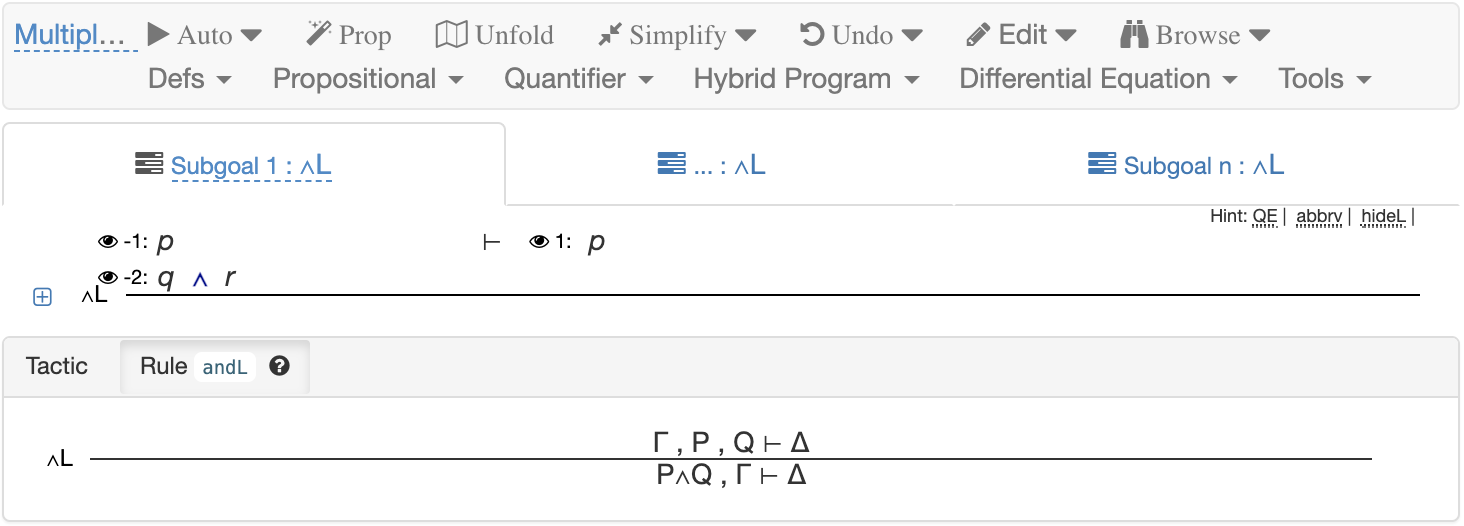}
};
\node[anchor=east] at ($(ui.north west)+(0,-0.6)$) {\footnotesize Tactics menu \circled{\bf A}};
\node[anchor=east] at ($(ui.west)+(0,0.3)$) {\footnotesize Deduction \circled{\bf B}};
\node[anchor=east] at ($(ui.south west)+(0,0.8)$) {\footnotesize Explanation \circled{\bf C}};
\node[anchor=west] at ($(ui.north west)+(0.2,-1.6)$) {\footnotesize\circled{\bf D}};
\end{tikzpicture}
\caption{User interface displays the open subgoals of a proof.
Tactics menu~$\tiny\circled{\bf A}$ lists proof tactics and automation, deduction view~$\tiny\circled{\bf B}$ lists the open subgoals (one tab per subgoal), explanation~$\tiny\circled{\bf C}$ illustrates the last applied proof step and lists the recorded proof history as a tactic.
Subgoal 1 is selected $\tiny\circled{D}$ and its sequent $\lsequent{p,q \land r}{p}$ is displayed.
}
\label{fig:proofstate-ui}
\end{subfigure}
\caption{Proof state data structure and rendering on the user interface}
\end{figure}

The \KeYmaeraX prover core represents proof state as derived rules called \texttt{Provables}, which list the conclusion to prove and the open subgoals, as illustrated in \rref{fig:proofstate-provable}.
Conclusion and subgoals are each represented with a sequent of the form $\lsequent{\Gamma}{\Delta}$: assumptions are in $\Gamma$, $\Delta$ lists the alternatives to prove.
The meaning of sequent $\lsequent{\Gamma}{\Delta}$ is that of \dL formula $\bigwedge_{p \in \Gamma}p \limply \bigvee_{q\in\Delta}q$.
Validity of the subgoals justifies validity of the conclusion; a proof is closed when there are no more open subgoals.
The user interface of \KeYmaeraX in \rref{fig:proofstate-ui} displays proof state in its deduction view $\tiny\circled{\bf B}$, provides automation and tactics $\tiny\circled{\bf A}$ to progress in the proof, and lists proof step explanations $\tiny\circled{\bf C}$ as well as a tactic summarizing the recorded proof history.
Proofs can be started from an initial conjecture, as well as from a (partial) tactic that advances the proof state according to the tactic steps and displays the remaining proof goals (further manual steps are then recorded and extend the provided tactic).
The deduction view strives for close mnemonic similarity to text books \cite{DBLP:conf/fide/MitschP16,DBLP:series/lncs/MitschP20} while maximizing screen estate use (it displays open subgoals in tabs to utilize the full screen width for each subgoal).
In principle, the user interface could use typesetting libraries such as MathJax, to resemble textbook appearance even more closely, but such attempts were abandoned for rendering performance reasons.

The \KeYmaeraX core is stateless, it does not keep track of proof state.
Instead, tactics and proof management outside the core keep track of \texttt{Provables} and instruct the core to apply operations on \texttt{Provables} to transform proof state, see \cite{DBLP:series/lncs/MitschP20} for a description of how tactics combine axioms and a comparison to alternative implementation approaches.
Major core operations are to
\begin{itemize}
\item create a \texttt{Provable}, which is allowed only from a small number of sources, the most important ones are \(\begin{aligned}\linfer
{\lsequent{\Gamma}{\Delta}}
{\lsequent{\Gamma}{\Delta}}
&&,\quad
\linfer[qear]
{\lclose}
{\lsequent{\Gamma}{\Delta}}
&&,\quad
\linfer
{\lclose}
{\lsequent{}{\dL~\text{axiom}}}\end{aligned}\)
(from left to right: starting a proof by justifying the conjecture from itself, real arithmetic facts, and \dL axioms);
\item apply another \texttt{Provable}, whose conjecture matches a subgoal syntactically to replace the existing subgoal with the subgoals of the other \texttt{Provable}, which we exploit for applying lemmas;
\item apply uniform substitution to replace predicate symbols with formulas, function symbols with terms, and program symbols with hybrid programs, which is useful to support definitions.
\end{itemize}

\begin{figure}[b!]
\begin{subfigure}[b]{\textwidth}
\centering
\input{fig/simpleproof.tex}
\caption{A simple proof instructing the core to create a new \texttt{Provable}, apply rule $\irref{implyr}$ in step~$\tiny\bcircled{1}$, and apply $\irref{qear}$ in step~$\tiny\bcircled{2}$, which obtains a \texttt{Provable} from a trusted solver and applies it to the remaining subgoal to close the proof.}
\label{fig:coreproof-internal}
\end{subfigure}
\begin{subfigure}[b]{\textwidth}
\centering
\begin{tikzpicture}
\node[draw,dashed] (implyr) {
\includegraphics[width=.77\textwidth]{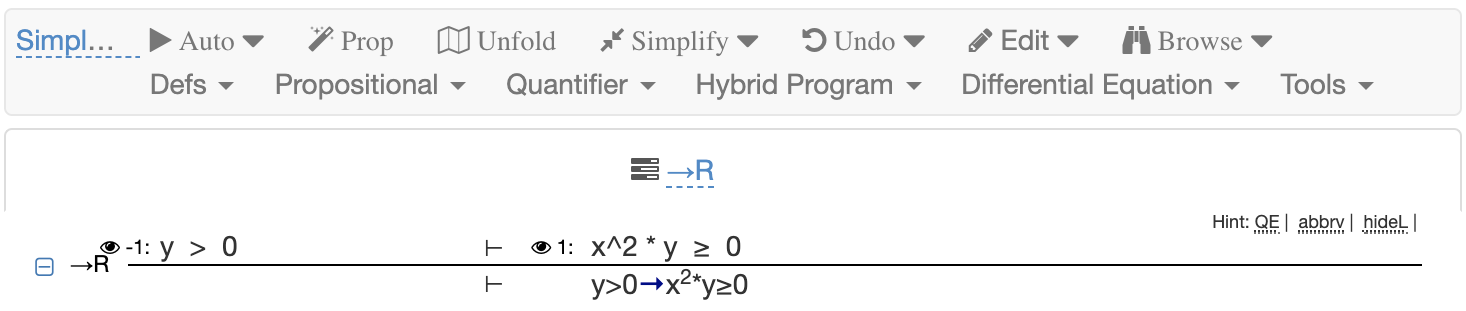}
};
\node[below=-0.7cm of ui,draw,dashed] (qe) {
\includegraphics[width=.77\textwidth]{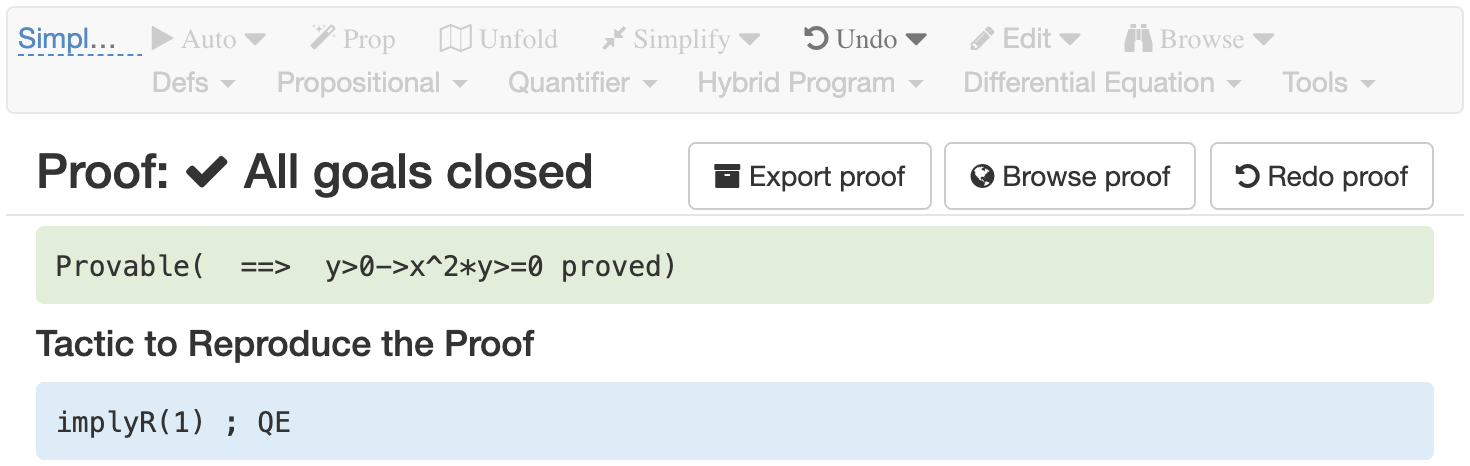}
};
\node[anchor=east] at ($(ui.west)+(0,0)$) {\footnotesize Step~\irref{implyr}~\bcircled{\bf 1}};
\node[draw,circle,very thick] (qehint) at ($(implyr.north east)+(-1.8,-2)$) {};
\draw[-latex] (qehint) -- node[right,pos=0.75,fill=white,xshift=0.5cm] {apply hint $\irref{qear}~\bcircled{2}$} (qe);
\node[anchor=east] at ($(qe.west)+(0,-0.3)$) {\footnotesize ASCII \texttt{Provable}~\circled{\bf A}};
\node[anchor=east] at ($(qe.west)+(0,-1.6)$) {\footnotesize Tactic~\circled{\bf B}};
\end{tikzpicture}
\caption{Presentation of step $\irref{implyr}~\tiny\bcircled{1}$ on the user interface as a sequent proof.
The final \texttt{Provable} is printed directly from the \KeYmaeraX core in ASCII syntax, the tactic below lists the steps to reproduce the proof from the conjecture.
}
\label{fig:coreproof-ui}
\end{subfigure}
\caption{Steps in a proof in the core vs. presentation on the user interface}
\label{fig:coreproof}
\end{figure}
\noindent A typical proof, illustrated in \rref{fig:coreproof-internal}, retrieves an initial \texttt{Provable} from the \KeYmaeraX core and then proceeds by handing back the \texttt{Provable} to the core together with a proof rule to retrieve a follow-up \texttt{Provable}.
This process is repeated until all subgoals are either reduced to \dL axioms or valid formulas in real arithmetic, so no more subgoals remain.
At any point in this process can proof state be stored and used later as a lemma (even in other proofs).
This entire process is hidden from the user, who instead is presented the sequent proof in \rref{fig:coreproof-ui}.

\KeYmaeraX proofs appeal to uniform substitution from \dL axioms \cite{DBLP:journals/jar/Platzer17}: for example, the test axiom $\dibox{\ptest{q}}p \lbisubjunct (q \limply p)$, which is an ordinary \dL formula, together with uniform substitution $\sigma=\usubstlist{\usubstmod{q}{x>0},\usubstmod{p}{\dbox{\pevolve{\D{x}=-x}}x\geq 0}}$ can be used to obtain a concrete instance of this axiom during a proof as follows:

\irlabel{US|US}

\[
\linfer[US]
{\dbox{\ptest{q}}p \lbisubjunct (q \limply p)}
{\dbox{\ptest{x>0}}\dbox{\pevolve{\D{x}=-x}}x\geq 0 \lbisubjunct (x>0 \limply \dbox{\pevolve{\D{x}=-x}}x\geq 0)} \enspace .
\]

\noindent Uniform substitution is mainly used as a mechanism to instantiate axioms soundly, but through \cite[Thm. 27]{DBLP:journals/jar/Platzer17} it is also useful to replace symbols in entire \texttt{Provables} soundly, as illustrated in \rref{fig:provablesubst}.
\begin{figure}
\centering
\begin{tikzpicture}
\node[text width=4cm,draw,dashed] (conjecture) {
  \begin{minipage}{\textwidth}
  \begin{sequentdeduction}[array]
  \linfer
  {\lsequent{p}{\dibox{a}{q}}
  !\lsequent{p}{\dibox{b}{q}}
  }
  {\lsequent{}{p \limply \dibox{\pchoice{a}{b}}{q}}}
  \end{sequentdeduction}
  \end{minipage}
};
\node[right=1.3cm of conjecture,text width=10cm,draw,dashed] (substituted) {
  \begin{minipage}{\textwidth}
  \begin{sequentdeduction}[array]
  \linfer
  {\lsequent{x>0}{\dibox{\pevolve{\D{x}=-x}}x>0}
  !\lsequent{x>0}{\dibox{\prepeat{(\humod{x}{\frac{x}{2}})}}x>0}
  }
  {\lsequent{}{x>0 \limply \dibox{\pchoice{\pevolve{\D{x}=-x}}{\prepeat{(\humod{x}{\frac{x}{2}})}}}x>0}}
  \end{sequentdeduction}
  \end{minipage}
};
\draw[-latex,solid,black] (conjecture) -- node[above] {\irref{US}~\bcircled{1}} (substituted);
\end{tikzpicture}
\caption{Uniform substitution $\sigma=\usubstlist{\usubstmod{p}{x>0},\usubstmod{a}{\pevolve{\D{x}=-x}},\usubstmod{b}{\prepeat{(\humod{x}{\frac{x}{2}})}},\usubstmod{q}{x>0}}$ on an entire \texttt{Provable} has uniform effect across subgoals and conclusion \cite[Thm. 27]{DBLP:journals/jar/Platzer17}.}
\label{fig:provablesubst}
\end{figure}
In this paper, we are going to exploit its application to entire \texttt{Provables} in order to implement proof features such as expanding definitions during a proof and an extended lemma mechanism that is able to bridge syntactic differences between the lemma conclusion and its application target.

\section{Implicit and Explicit Proof Management by Example}
\label{sec:proofmanagementexample}

\begin{figure}[b!]
\begin{subfigure}[b]{.52\textwidth}
\centering
\begin{lstlisting}[language=KeYmaeraX]
Definitions
  Bool A(Real x) <-> x=2;
  Bool S(Real x) <-> x>=0;
  HP ctrl ::= { if (S(x)) x:=2*x; }; /* ?S(x);x:=2*x; ++ ?!S(x);?true; */
  HP ode  ::= { {x'=-x} };
End.

ProgramVariables Real x; End.

Problem  A(x) -> [{ctrl;ode;}*]S(x) End.
\end{lstlisting}
\caption{Definitions in the \KeYmaeraX input syntax}
\label{fig:defs-inputfile}
\end{subfigure}
\quad
\begin{subfigure}[b]{.45\textwidth}
\centering
\includegraphics[width=\columnwidth]{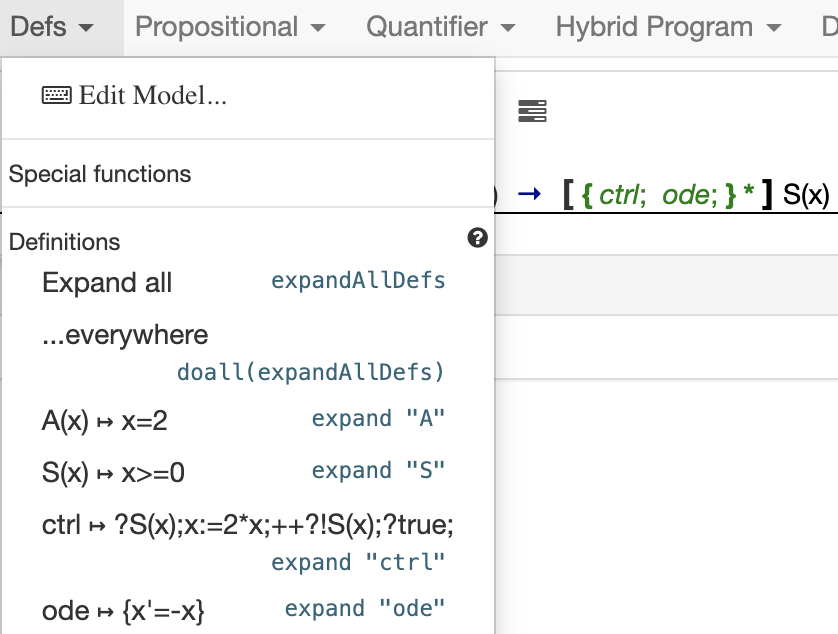}
\caption{Definitions menu}
\label{fig:defs-menu}
\end{subfigure}
\caption{Predicate and program definitions in \KeYmaeraX}
\label{fig:defs}
\end{figure}

The main motivation for proof management is to allow users expand definitions and structure proofs at their discretion, as well as to enable future automated definition expansion and contraction~\cite{DBLP:journals/jar/Wos87c}.
For example, consider the \KeYmaeraX input file in \rref{fig:defs-inputfile} that uses predicate definitions $A(x) \equiv x=2$ to capture assumptions about initial values of $x$ and $S(x) \equiv x \geq 0$ to describe the desired safety property, as well as program definitions $\textit{ctrl}$, which doubles the value of any non-negative $x$, and the differential equation $\textit{ode}$, which models exponential decay.
The definitions populate the ``Defs'' menu in \KeYmaeraX that allows users to expand definitions collectively (\texttt{expandAllDefs}) or selectively (e.g., \texttt{expand "ctrl"}) during a proof, see \rref{fig:defs-menu}.
The menu automatically adjusts to the symbols of the currently selected subgoal (in the background in \rref{fig:defs-menu}).

As a safety question example, we want to answer whether repeated execution of $\textit{ctrl};\textit{ode}$ keeps the value of $x$ non-negative when started at $x=2$.
In the proof, we want control over when to expand definitions, and we want to structure the proof into a main theorem and supporting lemmas.
\rref{fig:expanddefs} illustrates the proof steps.

\irlabel{expand|expand}
\irlabel{fold|fold}
\irlabel{ode|ode}
\irlabel{ids|id using S(x)}
\irlabel{by|by}
\irlabel{auto|auto}
\irlabel{MR|MR}

\begin{figure}[htb]
\begin{subfigure}[b]{.45\textwidth}
\begin{sequentdeduction}[array]
\linfer[ode]
{\lclose}
{
\linfer[testb+implyr]
{\lsequent{x \geq 0}{\dibox{\pevolve{\D{x}=-x}}x \geq 0}}
{
\lsequent{}{\dibox{\ptest{x \geq 0}}\dibox{\pevolve{\D{x}=-x}}x \geq 0}
}
}
\end{sequentdeduction}
\caption{Exponential decay lemma}
\label{fig:expanddefs-decaylemma}
\end{subfigure}
\quad
\begin{subfigure}[b]{.45\textwidth}
\begin{sequentdeduction}[array]
\linfer[ids]
{\lclose}
{
\linfer[notl]
{\lsequent{S(x)}{P,S(x)}}
{
\linfer[testb+implyr]
{\lsequent{S(x),\neg S(x)}{P}}
{
\linfer[implyr]
{\lsequent{S(x)}{\dibox{\ptest{\neg S(x)}}P}}
{
\lsequent{}{S(x) \limply \dibox{\ptest{\neg S(x)}}P}
}
}
}
}
\end{sequentdeduction}
\caption{Unsatisfied control guard lemma}
\label{fig:expanddefs-lemma}
\end{subfigure}
\begin{subfigure}[b]{\textwidth}
\vspace{\baselineskip}
\begin{sequentdeduction}[array]
\linfer[choiceb+andr]
{
\linfer[MR]
{
\linfer[by]
{\rref{fig:expanddefs-decaylemma}}
{
\lsequent{S(x)}{\dibox{\textit{ode}}{S(x)}}
}
!
\linfer[auto]
{\lclose}
{\lsequent{S(x)}{\dbox{\ptest{S(x)};\humod{x}{2x}}S(x)}
}
}
{
\lsequent{S(x)}{\dibox{\ptest{S(x)};\humod{x}{2x}}\dibox{\textit{ode}}{S(x)}}
}
!
\linfer[by]
{\rref{fig:expanddefs-lemma}}
{
\lsequent{S(x)}{\dibox{\ptest{\neg S(x)}}\dibox{\textit{ode}}{S(x)}}
}
}
{
\linfer[expand]
{\lsequent{S(x)}{\dibox{\pchoice{\ptest{S(x)};\humod{x}{2x}}{\ptest{\neg S(x)}}}\dibox{\textit{ode}}{S(x)}}}
{
\linfer[composeb]
{\lsequent{S(x)}{\dibox{\textit{ctrl}}\dibox{\textit{ode}}S(x)}}
{
\lsequent{S(x)}{\dibox{\textit{ctrl};\textit{ode}}S(x)}
}
}
}
\end{sequentdeduction}
\caption{Induction step lemma, appeals to exponential decay lemma and unsatisfied guard lemma}
\label{fig:expanddefs-step}
\end{subfigure}
\begin{subfigure}[b]{\textwidth}
\begin{sequentdeduction}[array]
\linfer[loop]
{
\linfer[qear]
{\lclose}
{
\linfer[expand]
{\lsequent{x=2}{x \geq 0}}
{\lsequent{A(x)}{S(x)}
}
}
!
\linfer[by]
{\rref{fig:expanddefs-step}}
{
\lsequent{S(x)}{\dibox{\textit{ctrl};\textit{ode}}S(x)}
}
!
\linfer[id]
{\lclose}
{\lsequent{S(x)}{S(x)}
}
}
{
\linfer[implyr]
{\lsequent{A(x)}{\dibox{\prepeat{(\textit{ctrl};\textit{ode})}}S(x)}}
{
\lsequent{}{A(x) \limply \dibox{\prepeat{(\textit{ctrl};\textit{ode})}}S(x)}
}
}
\end{sequentdeduction}
\caption{Main theorem proves loop induction base case and use case, appeals to \rref{fig:expanddefs-step} for the induction step.}
\label{fig:expanddefs-main}
\end{subfigure}
\caption{The substitutions collected during step~\texttt{expand} and step~\texttt{by} are $\sigma=\usubstlist{\usubstmod{A(\cdot)}{\cdot=2},\usubstmod{S(\cdot)}{\cdot \geq 0},\usubstmod{ctrl}{\pchoice{\ptest{S(x)}}{\ptest{\neg S(x)}}},\usubstmod{ode}{\pevolve{\D{x}=-x}},\usubstmod{P}{\dibox{ode}S(x)}}$, so the proof shows validity of the concrete formula $x=2 \limply \dibox{\prepeat{\bigl((\pchoice{\ptest{x \geq 0}}{\ptest{\neg x \geq 0}});\{\D{x}=-x\}\bigr)}}x \geq 0$.}
\label{fig:expanddefs}
\end{figure}

The proof proceeds from the initial conjecture $\lsequent{A(x)}{\dibox{\prepeat{(\textit{ctrl};\textit{ode})}}S(x)}$ bottom-to-top, with proof step justifications annotated to the left of the horizontal bars.
Validity transfers top-to-bottom, so validity of the sequents (subgoals) above a horizontal bar justifies validity of the conclusion below the horizontal bar.
The first step~\irref{implyr} makes the left-hand side of the implication available as assumptions $A(x)$.
Next, step~\irref{loop} induction splits the proof into three subgoals: the base case $\lsequent{A(x)}{S(x)}$, which closes by real arithmetic~\irref{qear} after expanding the definitions, the use case $\lsequent{S(x)}{S(x)}$ that is trivially true by step~\irref{id}, and the induction step.
The induction step in \rref{fig:expanddefs-step} first addresses the sequential composition with step~\irref{composeb} to isolate $\textit{ctrl}$ from $\textit{ode}$, then expands $\textit{ctrl}$ to split into its two cases:
\begin{inparaenum}[(i)]
\item on the left branch, the condition of $\text{if}~(S(x))$ is true (represented with $\ptest{S(x)}$) and preserved by the program $\ptest{S(x)};\humod{x}{2}$ as witnessed by a monotonicity step~\irref{MR} and, thus, the exponential decay lemma applies;
\item on the right branch with its leading test $\ptest{\neg S(x)}$ the unsatisfied control guard lemma applies.
\end{inparaenum}

The main proof management features used in the proof are step~\irref{expand} to expand definitions, step~\irref{by} to apply a lemma, and \texttt{using} to temporarily restrict reasoning to certain formulas.
To users, the proof in \rref{fig:expanddefs} appears as if they were working on a single \texttt{Provable} and the proof steps were combined immediately.
Doing so, however, would require extensive changes to the soundness-critical core and violate the local nature of its reasoning.
Behind the scenes, this proof therefore requires a shift from operating on a single \texttt{Provable} to keeping track of loosely connected sub-proofs outside the prover core; these sub-proofs fit together only after applying the substitutions collected during the proof.
In the following sections, we provide details on explicit proof management that structures proofs into lemmas and implicit proof management that delays merging \texttt{Provables} and applying uniform substitutions.

\section{Explicit Proof Management with Lemmas}
\label{sec:lemmas}

The \KeYmaeraX input format allows explicit proof management in the input format for users to structure their problem descriptions into lemmas that are shared between proofs, and theorems, which appeal to lemmas to show some of their subgoals.
For example, the ``Exponential decay'' and ``Unsatisfied control guard'' lemmas from \rref{fig:expanddefs-decaylemma} and \rref{fig:expanddefs-lemma} are expressed in the \KeYmaeraX ASCII input syntax below, recorded from the steps of the interactive proofs in \rref{fig:expanddefs-decaylemma} and \rref{fig:expanddefs-lemma}.
Optional ``/'' in lemma names structure the lemmas into folders on both the user interface and the file system.
\begin{lstlisting}[language=KeYmaeraX]
Lemma "FIDE21/Exponential decay"
  ProgramVariables Real x; End.
  Problem x>=0 -> [{x'=-x}]x>=0 End.
  Tactic "Recorded" implyR('R=="x>=0 -> [{x'=-x}]x>=0"); ODE('R=="[{x'=-x}]x>=0") End.
  Tactic "Automated proof" autoClose End.
End.
Lemma "FIDE21/Unsatisfied control guard"
  Definitions		 /* constants, functions, properties, programs */
    Bool S(Real x);
    Bool P(Real x);
  End.
  ProgramVariables Real x; End.       /* variables */
  Problem	S(x) -> [?!S(x);]P(x) End.  /* specification in dL */
  Tactic "Interactive proof"
    implyR('R=="S(x) -> [?!S(x);]P(x)");
    testb('R=="[?!S(x);]P(x)");
    implyR('R=="!S(x) -> P(x)");
    notL('L=="!S(x)");
    id using "S(x)"
  End.
  Tactic "Automated proof" autoClose End.
End.
\end{lstlisting}
\label{lst:lemmaFig1a}

The ``Induction step'' lemma below uses the earlier two lemmas in its proof.
It follows the steps in \rref{fig:expanddefs-step} largely verbatim, but the specific lemma application steps are worth noting.
Applying the ``Exponential decay'' lemma is straightforward by \irref{auto}, since it uses $S(x)$ and $\textit{ode}$ in their expanded form as the only difference between the subgoal and the lemma conclusion.
The ``Unsatisfied control guard'' lemma, however, introduces a new predicate symbol $P(x)$, which is neither present in the induction step nor in the original conjecture.
We, therefore, use substitution $\sigma=\usubstlist{\usubstmod{P(x)}{\dbox{\humod{x}{x}}\dbox{\ptest{\ltrue}}\dbox{\pevolve{\D{x}=-x}}S(x)}}$ to tell the lemma application mechanism how to resolve $P(x)$.\footnote{The leading self-assignment $\humod{x}{x}$ is a necessary technicality to make variable $x$ must-bound because the differential equation $\D{x}=-x$ may run for duration $0$.}
\begin{lstlisting}[language=KeYmaeraX]
Lemma "FIDE21/Induction step"
  Definitions
    Bool S(Real x) <-> x>=0;
    HP ctrl ::= { if (S(x)) { x:=2*x; } };
    HP ode ::= { {x'=-x} };
  End.
  ProgramVariables Real x; End.
  Problem S(x) -> [ctrl;ode;]S(x) End.
  Tactic "Proof induction step"
    implyR('R=="S(x)->[ctrl;ode;]S(x)");
    composeb('R=="[ctrl;ode;]S(x)");
    expand "ctrl";
    choiceb('R=="[?S(x);x:=2*x;++?!S(x);?true;][ode;]S(x)");
    andR('R=="[?S(x);x:=2*x;][ode;]S(x) & [?!S(x);?true;][ode;]S(x)"); <(
      "[?S(x);x:=2*x;][ode;]S(x)":
        MR("S(x)", 'R=="[?S(x);x:=2*x;][ode;]S(x)"); <(
          "Use Q->P": expand "S"; autoClose,
          "Show [a]Q": useLemma("FIDE21/Exponential decay", "US({`S(x)~>x>=0 :: ode;~>{x'=-x} :: nil`});unfold;id")
        ),
      "[?!S(x);?true;][ode;]S(x)":
        composeb('R=="[?!S(x);?true;][ode{|^@|};]S(x)");
        useLemma("FIDE21/Unsatisfied control guard", "US({`P(x)~>[x:=x;][?true;][{x'=-x}]S(x) :: nil`});unfold;id")
    )
  End.
End.
\end{lstlisting}

The main theorem of \rref{fig:expanddefs-main} is expressed in \KeYmaeraX ASCII syntax below.
Its proof uses the ``Induction step'' lemma in a straightforward way.
\begin{lstlisting}[language=KeYmaeraX]
Theorem "FIDE21/Combine lemmas"
  Definitions
    Bool A(Real x) <-> x=2;
    Bool S(Real x) <-> x>=0;
    HP ctrl ::= { if (S(x)) { x:=2*x; } };
    HP ode ::= { {x'=-x} };
  End.
  ProgramVariables Real x; End.
  Problem A(x) -> [{ctrl;ode;}*]S(x) End.
  Tactic "Interactive proof"
    implyR('R=="A(x)->[{ctrl;ode;}*]S(x)");
    loop("S(x)", 'R=="[{ctrl;ode;}*]S(x)"); <(
      "Init": expandAllDefs; QE,
      "Post": id,
      "Step": expandAllDefs; useLemma("FIDE21/Induction step", "prop")
    )
  End.
End.
\end{lstlisting}
\label{lst:theoremFig1}

\begin{figure}[htb]
\centering
\includegraphics[width=.7\textwidth]{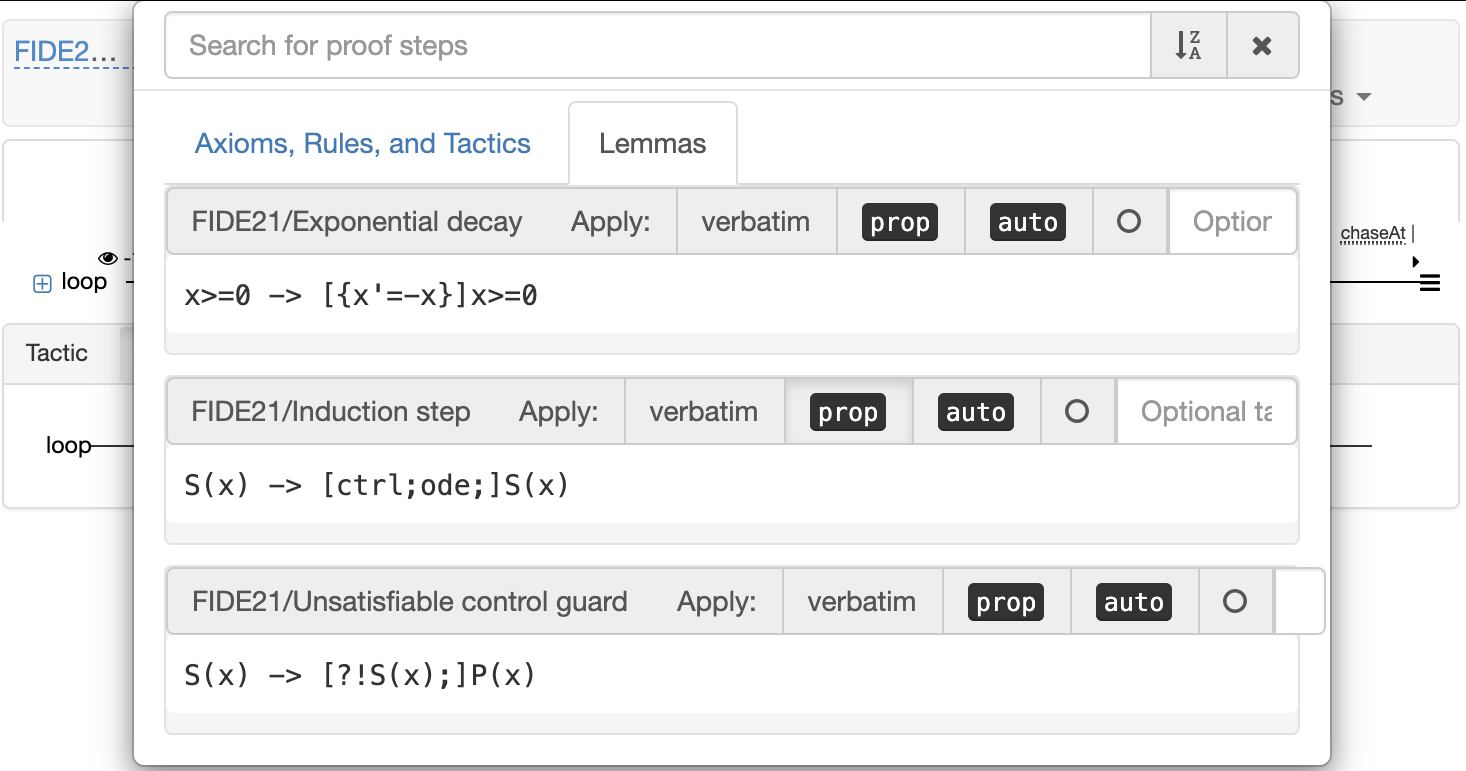}
\caption{Applying a lemma from the lemma search dialog in \KeYmaeraX}
\label{fig:applylemma}
\end{figure}

Structuring proofs into lemmas and theorems need not necessarily be done when creating the input file.
As an alternative, the \KeYmaeraX user interface allows users to start lemmas from any proof state; lemma proofs remain linked from the tabs representing open subgoals in the main proof until finished.
Other pre-existing lemmas can be searched and applied from the user interface as in \rref{fig:applylemma}.
Techniques for implicit proof management and delayed substitution (used in the proofs above) are discussed next.

\section{Implicit Proof Management with Delayed Substitution}
\label{sec:prooftechniques}

In this section, we discuss the fundamental proof management technique of delayed proof composition and delayed uniform substitution, and then devise several applications of it for enabling parametric proofs and delayed modeling, and for temporary sub-proofs focusing on some select aspects of a subgoal.

\subsection{Delayed Proof Composition and Delayed Uniform Substitution}
As illustrated in \rref{sec:preliminaries}, the \KeYmaeraX core creates and modifies proof state without keeping track of the steps of the proof.
\KeYmaeraX records proof steps outside the core with a separate \texttt{Provable} per proof step.
This trace of \texttt{Provables} not only is the basis for rendering and navigating the sequent calculus proof on the user interface, but also gives us freedom to choose when to combine \texttt{Provables}.
Instead of combining provables and applying uniform substitutions immediately at every step in the proof, those separate \texttt{Provables} are combined to a single proof once all the steps are finished.
The advantage of delayed merging is that we can postpone the uniform effect \cite[Thm. 27]{DBLP:journals/jar/Platzer17} of uniform substitution across subgoals of a \texttt{Provable}.
Without delayed merging, symbols that are expanded on one branch would immediately be expanded uniformly across all other branches of the proof, even if those other branches would prefer to continue using symbols in their unexpanded form.
The different points of expanding symbols are then reconciled in the final proof checking pass that combines \texttt{Provables}: uniform substitutions that originate from explicit user interactions (e.g., from expanding definitions) are combined with substitutions found through unification.

\subsection{Parametric Proofs and Delayed Modeling}

Theorem proving is not merely a tool to just obtain a correctness proof about an already correct system; it is a tool to explore and understand a system in all its subtleties and with all its corner cases thoroughly, to discover properties of the system that are not or only partially known, and to discover and fix correctness bugs in the process.
We, therefore, frequently want to make progress in a proof without yet committing to specific inputs or even without supplying a finished model and/or conjecture.
For example, we may want to analyze a loop, but do not yet know a concrete loop invariant that we could use in the proof.
An obvious technique is to use loop unrolling to debug the behavior of the loop body in an attempt to manually identify a loop invariant candidate.
However, this is usually not a suitable technique to prove safety of a loop, and  so requires duplicate proof effort once a loop invariant candidate is identified through debugging (and perhaps several rounds of alternating debugging and proof attempts).

\paragraph{Parametric Proofs} A powerful alternative technique are parametric proofs~\cite{DBLP:journals/jar/Platzer17} to advance in a proof without committing to concrete inputs early in a proof.
\begin{figure}[htb]
\begin{subfigure}[b]{\textwidth}
\begin{sequentdeduction}[array]
\linfer[loop]
{
\linfer[qear]
{\lclose}
{
\linfer[expand]
{\lsequent{x=2}{x \geq 1}}
{
\lsequent{x=2}{J(x)}
}
}
!
\linfer[qear]
{\lclose}
{
\linfer[expand]
{\lsequent{x \geq 1}{1+\frac{x-1}{2} \geq 1}}
{
\linfer[assignb]
{\lsequent{J(x)}{J(1+\frac{x-1}{2})}}
{\lsequent{J(x)}{\dibox{\humod{x}{1+\frac{x-1}{2}}}J(x)}
}
}
}
!
\linfer[qear]
{\lclose}
{
\linfer[expand]
{\lsequent{x \geq 1}{x \geq -1}}
{
\lsequent{J(x)}{x \geq -1}
}
}
}
{\lsequent{x=2}{\dibox{\prepeat{(\humod{x}{1+\frac{x-1}{2}})}}x \geq -1}}
\end{sequentdeduction}
\caption{Parametric sequent proof}
\end{subfigure}
\begin{subfigure}[b]{\textwidth}
\centering
\vspace{.5\baselineskip}
\begin{tikzpicture}
\node (start) {\includegraphics[width=.85\textwidth]{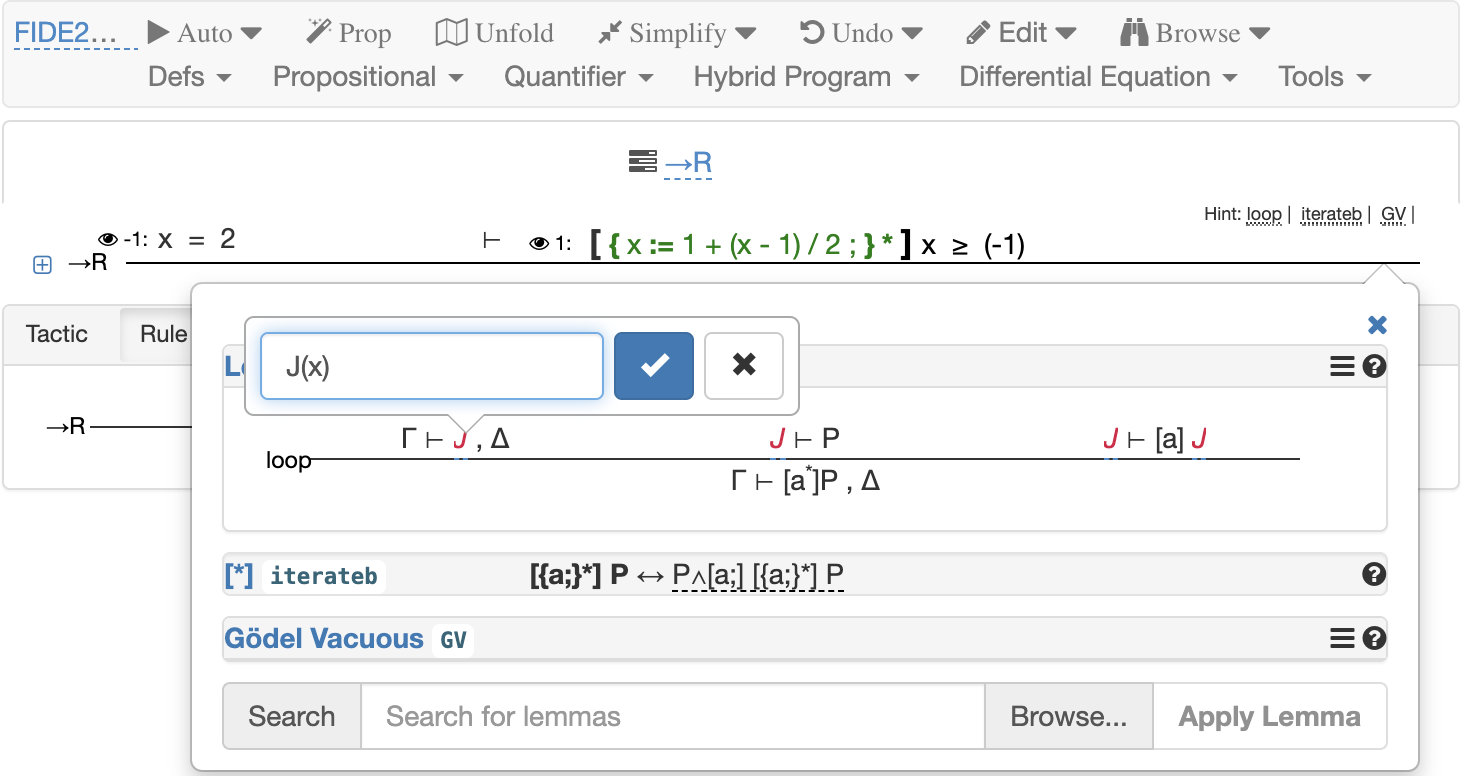}};
\node[draw,dashed,text width=0.8cm,minimum height=.5cm] at (-5,2.9) (menuanchor) {};
\node[below left=0cm and -1.8cm of start.west,anchor=east,draw,dashed] (menu) {
\includegraphics[width=4cm]{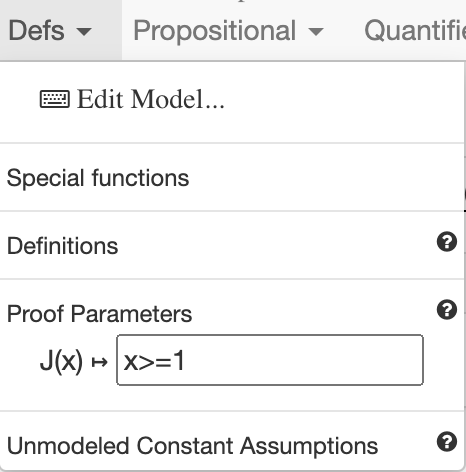}
};
\draw[-,dashed] (menuanchor.south west) -- (menu.north west);
\draw[-,dashed] (menuanchor.south east) -- (menu.north east);
\end{tikzpicture}
\caption{Starting a parametric proof in \KeYmaeraX by using an uninterpreted predicate symbol as tactic input; the proof parameter is then definable from a text field in the ``Defs'' menu.}
\end{subfigure}
\caption{Parametric loop induction with abstract invariant $J(x)$.
The substitution $\sigma=\{J(\cdot)\mapsto \cdot \geq 1\}$ supplies a concrete loop invariant to close all three branches of the proof, but any other substitution $\sigma=\{J(\cdot)\mapsto \cdot \sim y\}$ with $\sim\, \in \{\geq,>\}$ and $y \in [-1,1]$ would work as well.
}
\label{fig:parametricloopproof}
\end{figure}
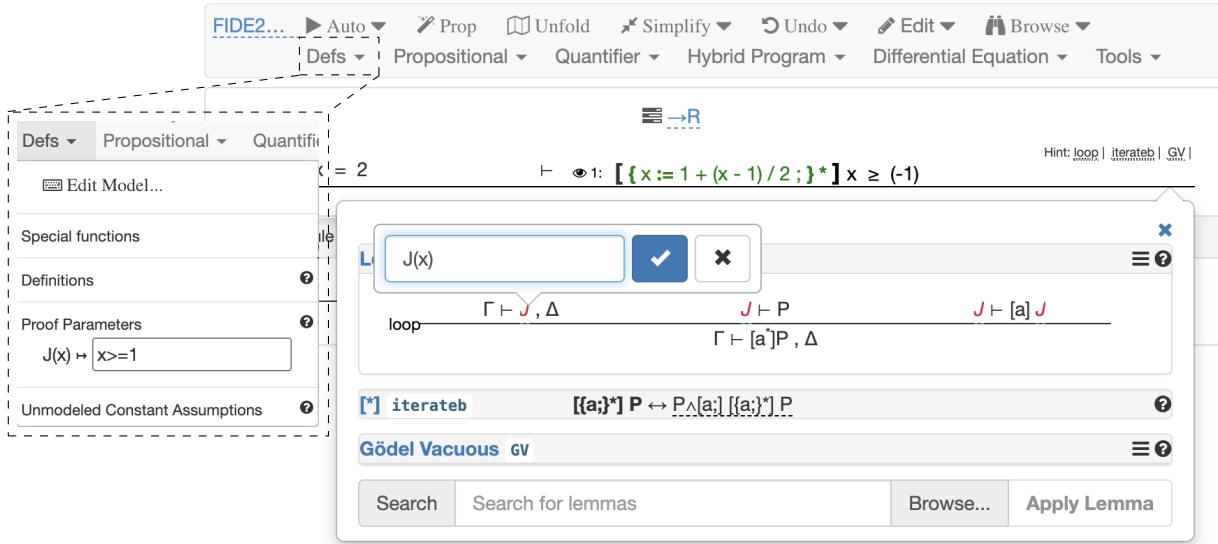
Parametric proofs allow users to proceed with abstract terms or formulas whose concrete shape is discovered later during the proof.
Delayed merging of \texttt{Provables} and delayed uniform substitution allow users to supply concrete terms, formulas, and programs for function symbols, predicate symbols, and program symbols at any point in the proof, which then get automatically applied to all prior proof steps upon composition of the final \texttt{Provable}.

In \rref{fig:parametricloopproof}, step~\irref{loop} uses an uninterpreted predicate symbol $J(x)$ instead of a concrete formula as a loop invariant.
That way, we can advance the proof on all three branches until we find that $J(x)$ simultaneously has to fit $\lsequent{x=2}{J(x)}$ in the induction base case, $\lsequent{J(x)}{J(1+\frac{x-1}{2})}$ in the induction step, and $\lsequent{J(x)}{x \geq -1}$ in the induction use case.
At this point, we can experiment with different choices of $J(x)$ and, ultimately, settle for $\sigma=\usubstlist{\usubstmod{J(\cdot)}{\cdot \geq 1}}$.
Uniformly substituting this choice into the entire \texttt{Provable} concludes the proof by \irref{qear} on all branches.

\begin{figure}[htb]
\begin{subfigure}[b]{.4\textwidth}
\begin{sequentdeduction}[array]
\linfer[qear]
{\lclose}
{
\linfer[expand]
{\lsequent{x \geq 0, y \neq 0}{\frac{x}{y^2} \geq 0}}
{
\linfer[assignb]
{\lsequent{x \geq 0, p_\text{\faShoppingCart}}{\frac{x}{y^2} \geq 0}}
{\lsequent{x \geq 0, p_\text{\faShoppingCart}}{\dibox{\humod{x}{\frac{x}{y^2}}}x \geq 0}
}
}
}
\end{sequentdeduction}
\caption{Predicate $p_\text{\faShoppingCart}$ to supply assumptions during the proof.
The substitution $\sigma=\{p_\text{\faShoppingCart}\mapsto y\neq0\}$ results in the proof showing validity of the sequent $\lsequent{x\geq 0,y\neq 0}{\dibox{\prepeat{(\humod{x}{\frac{x}{y^2}})}}x\geq0}$.}
\end{subfigure}
\quad
\begin{subfigure}[b]{.55\textwidth}
\centering
\begin{tikzpicture}
\node (sequent) {
\includegraphics[width=\textwidth]{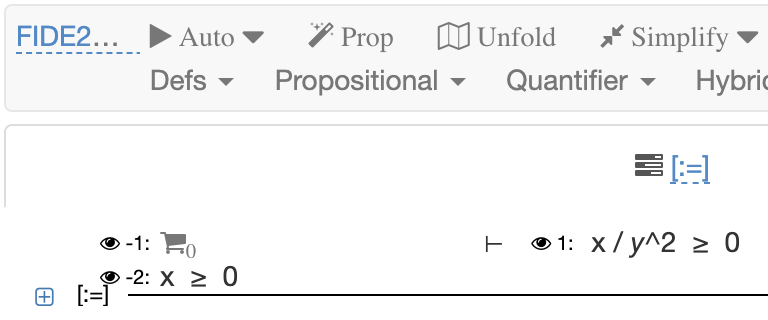}
};
\node[below left=1.5cm and 1.8cm of sequent.north,anchor=north west,draw,dashed] (menu) {
\includegraphics[width=4cm]{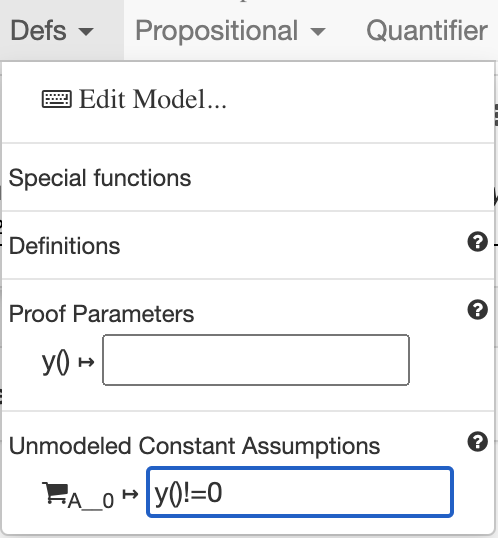}
};
\node[below right=0.8cm and 1.7cm of sequent.north west,text width=0.8cm,minimum height=0.5cm,draw,dashed] (menuanchor) {};
\draw[-,dashed] (menuanchor.south west) -- (menu.north west);
\draw[-,dashed] (menuanchor.south east) -- (menu.north east);
\end{tikzpicture}
\caption{Providing assumptions with the ``Defs'' menu}
\end{subfigure}
\caption{Delayed modeling by using uninterpreted predicate symbols in the input}
\label{fig:delayedassumptions}
\end{figure}
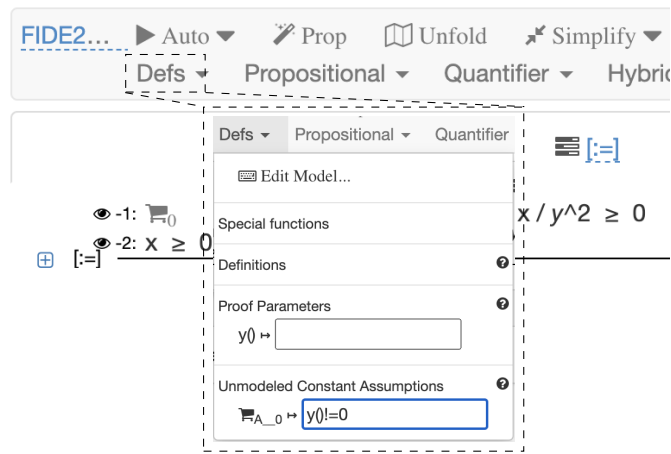

\paragraph{Delayed Modeling}
Delayed merging of \texttt{Provables} and delayed substitution are also helpful to address a common nuisance in proofs: missing assumptions about model parameters (e.g., to avoid division by zero) are easily forgotten and their absence becomes apparent often only rather late in a proof.
Using an arity 0 predicate symbol in the model allows users to supply missing assumptions during a proof as they are discovered, without requiring to redo the proof.
For example, the proof in \rref{fig:delayedassumptions} uses an arity 0 predicate symbol $p_\text{\faShoppingCart}$ that can be used to collect missing assumptions as they become apparent in the proof.
The effect of collecting assumptions during the proof is achieved by simply augmenting the concrete assumption with another fresh $p_{\text{\faShoppingCart}i}$.

Note that a proof parameter $J(x)$ cannot be used to introduce the missing assumption $y \neq 0$ because that fact was not even available in the original conjecture and therefore would not be provable in the base case of the induction proof.
We use $p_\text{\faShoppingCart}$ to allow limited fixing of model mistakes during the proof; in the proof in \rref{fig:delayedassumptions} it is important that $p_\text{\faShoppingCart}$ is an arity 0 predicate symbol whose free variables do not overlap with the variables bound in the loop, so that it stays available in the induction step of the proof.
The conjecture of \rref{fig:delayedassumptions} is expressed below in \KeYmaeraX ASCII input syntax.
\begin{lstlisting}[language=KeYmaeraX]
Problem A__0() -> x>=0 -> [x:=x/y()2;]x>=0 End.
\end{lstlisting}

\subsection{Temporary Implicit Sub-Proofs with Select Formulas}

Applying tactics, axioms, and proof rules has permanent effect on the proof state.
For example, weakening assumptions permanently removes formulas from the proof state; if weakening is done for convenience to focus on specific aspects of the proof, we cannot undo the effect of weakening when the hidden formulas become useful again later in the proof.
Not focusing, however, is not an option either, because the mere presence of additional assumptions and formulas may result in duplicate proof effort or intractable proofs (e.g., when applying real arithmetic decision procedures with non-trivial complexity).

\begin{figure}
\begin{subfigure}[b]{\textwidth}
\begin{sequentdeduction}[array]
\linfer[orl]
{
\linfer[qear]
{\lclose}
{
\lsequent{\textcolor{gray}{\text{\faEyeSlash}\,x<0 \lor x=0},y=x}{xy \leq y^2}
}
!
\linfer[qear]
{\lclose}
{
\linfer[weakenli]
{\lsequent{x<0 \lor x=0,y>0}{xy \leq y^2}}
{
\lsequent{\textcolor{gray}{\text{\faEyeSlash}\,x<0 \lor x=0},y>0}{xy \leq y^2}
}
}
}
{
\linfer[weakenl]
{\lsequent{\textcolor{gray}{\text{\faEyeSlash}\,x<0 \lor x=0},y=x \lor y>0}{xy \leq y^2}}
{\lsequent{x<0 \lor x=0,y=x \lor y>0}{xy \leq y^2}
}
}
\end{sequentdeduction}
\caption{Wanted effect of temporarily hiding a formula: ignore for some proof steps, re-introduce when needed}
\label{fig:temphide-ok}
\end{subfigure}
\begin{subfigure}[b]{.45\textwidth}
\begin{sequentdeduction}[array]
\linfer[id]
{\lclose}
{
\linfer[weakenli]
{\lsequent{x=0,x=1}{x=0}}
{
\linfer[unsoundassignb]
{\lsequent{\textcolor{gray}{\text{\faEyeSlash}\,x=0},x=1}{x=0}}
{
\linfer[weakenl]
{\lsequent{\textcolor{gray}{\text{\faEyeSlash}\,x=0}}{\dibox{\humod{x}{1}}x=0}}
{\lsequent{x=0}{\dibox{\humod{x}{1}}x=0}
}
}
}
}
\end{sequentdeduction}
\caption{Unsoundly ignoring temporarily hidden formula}
\label{fig:temphide-assignunsound}
\end{subfigure}
\qquad
\begin{subfigure}[b]{.5\textwidth}
\begin{sequentdeduction}[array]
\linfer[DUS]
{\lsequent{x_0=0,x=1}{x=0}}
{
\linfer[assignb]
{\lsequent{P_\text{\faEyeSlash}(x_0),x=1}{x=0}}
{
\linfer[DUS]
{\lsequent{P_\text{\faEyeSlash}(x)}{\dibox{\humod{x}{1}}x=0}}
{\lsequent{x=0}{\dibox{\humod{x}{1}}x=0}
}
}
}
\end{sequentdeduction}
\caption{Sub-proof with substitution $\sigma=\{P_\text{\faEyeSlash}(\cdot)\mapsto \cdot=0\}$}
\label{fig:temphide-assignok}
\end{subfigure}
\begin{subfigure}[b]{\textwidth}
\centering
\vspace{\baselineskip}
\includegraphics[width=8cm]{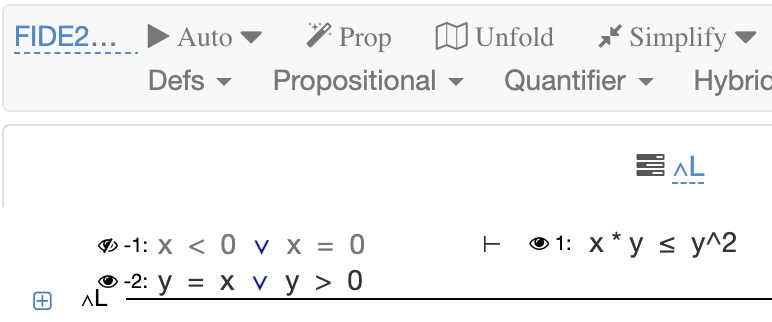}
\caption{Hiding formulas temporarily on the user interface}
\end{subfigure}
\caption{Sub-proof allows hiding explicit formula structure to temporarily focus tactic application on relevant formulas without sacrificing soundness.}
\label{fig:temphide}
\end{figure}

We want to keep temporary operations separate from the soundness-critical prover core, because their effect is not compatible with its local isolated reasoning principles.
It would be unsound to temporarily exclude formulas so that they are not affected by tactic applications.
\rref{fig:temphide-ok} illustrates an example with the wanted effect of temporarily hiding formulas, but \rref{fig:temphide-assignunsound} shows that care needs to be taken to not unsoundly exclude formulas temporarily from being affected by, e.g., the assignment axiom.
In order to fix \rref{fig:temphide-assignunsound}, the prover core would need to know which axiom or rule can soundly ignore temporarily hidden facts under what conditions.
With an implicit sub-proof as in \rref{fig:temphide-assignok} we can temporarily focus tactic application on some proof aspects without extending the \KeYmaeraX core or sacrificing soundness.
As an additional benefit, the abbreviations $P_\text{\faEyeSlash}(x)$ have simpler structure than the fully expanded formulas, which makes tactic applications less expensive even if they have to operate on $P_\text{\faEyeSlash}(x)$.

\irlabel{weakenli|W\leftrule i}
\irlabel{unsoundassignb|$\lightning$}

In tactics, temporarily focusing on a subset of the sequent formulas is supported with the notation \texttt{using}.
For example, the proof of \rref{fig:temphide-ok} is expressed as follows:
\begin{lstlisting}[language=Bellerophon]
(orL('L)*; <(QE, skip)) using "y=x|y>0 :: x*y<=y^2 :: nil"; QE
\end{lstlisting}
This script advances the proof fully on the left branch, but postpones the final \irref{qear} (\texttt{QE}) step of the right branch using \texttt{skip} until after the \texttt{using} block.

\section{Conclusion}
\label{sec:conclusion}

Uniform substitution in hybrid systems is a powerful technique for implementing hybrid systems theorem provers in an LCF-style approach, but it comes at the expense of proof convenience when sticking exclusively to core operations.
We illustrated how proof convenience can be regained with proof management features that are implemented on top of uniform substitution outside the soundness-critical core of a theorem prover, and we complement those features with modeling conventions and corresponding treatment in the user interface.

This approach of using modeling conventions and making proof steps implicit through other user interactions sits somewhat between auto-active verifiers and full interactive theorem proving.
Auto-active verifiers, such as Dafny \cite{DBLP:conf/icse/Leino04,DBLP:journals/corr/LeinoW14} and AutoProof \cite{DBLP:conf/tacas/TschannenFNP15} hide verification and interaction with the verification tool entirely behind annotations in the analyzed code.
Interactive theorem provers, such as Coq~\cite{DBLP:series/txtcs/BertotC04} and Isabelle/HOL~\cite{DBLP:books/sp/NipkowPW02}, primarily interact with users through scripts, such as structured proofs in Isabelle/Isar~\cite{DBLP:conf/types/Nipkow02} and  hide only little of the proof complexity behind other means of presentation even in advanced editors \cite{DBLP:conf/aisc/Wenzel12} or when proofs are found automatically, e.g., with Sledgehammer~\cite{DBLP:conf/cade/Paulson10}.
Many (hybrid systems) theorem provers (e.g., \cite{KeY2005,DBLP:conf/cade/PlatzerQ08,DBLP:conf/icfem/RenshawLP11,DBLP:conf/icfem/WangZZ15}) opt for implementing their proof calculus using axiom schemata or with trusted rules, which renders the features presented here soundness-critical.
Our proof management techniques, in contrast, provide proof convenience without sacrificing soundness.

For future work, we plan to automate unification steps in applying lemmas to bridge the syntactic differences between lemma conclusion and target subgoal, and seek to exploit uniform substitution for further proof techniques.

\bibliographystyle{eptcs}
\bibliography{refs,refs2}
\end{document}

%% file: fig/simpleproof.tex
\providecommand\circled[1]{\tikz[baseline=(char.base)]{
            \node[shape=circle,draw,inner sep=2pt] (char) {#1};}}
\providecommand\bcircled[1]{\tikz[baseline=(char.base)]{
            \node[shape=circle,draw,inner sep=2pt,fill=black,text=white] (char) {\textbf{#1}};}}

\begin{tikzpicture}
\node[text width=3cm,draw,dashed] (conjecture) {
  \begin{minipage}{\textwidth}
  \begin{sequentdeduction}[array]
  \linfer
  {\lsequent{}{y>0 \limply x^2y\geq0}}
  {\lsequent{}{y>0 \limply x^2y\geq0}}
  \end{sequentdeduction}
  \end{minipage}
};
\node[right=2cm of conjecture,text width=4cm,draw,dashed] (implyr) {
  \begin{minipage}{\textwidth}
  \begin{sequentdeduction}[array]
  \linfer
  {\lsequent{y>0}{x^2y\geq0}}
  {\lsequent{}{y>0 \limply x^2y\geq0}}
  \end{sequentdeduction}
  \end{minipage}
};
\node[below right=-0.2cm and 0.3cm of implyr,text width=2cm,draw,dashed] (qe) {
  \begin{minipage}{\textwidth}
  \tiny
  \begin{sequentdeduction}[array]
  \linfer[qear]
  {\lclose}
  {\lsequent{y>0}{x^2y\geq0}}
  \end{sequentdeduction}
  \end{minipage}
};
\node[right=3cm of implyr,text width=3cm,draw,dashed] (result) {
  \begin{minipage}{\textwidth}
  \begin{sequentdeduction}[array]
  \linfer
  {\lclose}
  {\lsequent{}{y>0 \limply x^2y\geq0}}
  \end{sequentdeduction}
  \end{minipage}
};
\draw[-latex,solid,black] (conjecture) -- node[above] {\irref{implyr}~\bcircled{1}} (implyr);
\draw[-latex,dashed,black] (implyr) -- node[above] {\irref{qear}~\bcircled{2}} (result);
\draw[-latex,solid,black] ($(implyr.east)+(0,-0.2)$) -| ($(qe.north)+(-0.2,0)$);
\draw[-latex,solid,black] ($(qe.north)+(0.2,0)$) |- node[right,pos=0.2] {} ($(result.west)+(0,-0.2)$);
\end{tikzpicture}

%% file: delayedsubst.bbl
\begin{thebibliography}{10}
\providecommand{\bibitemdeclare}[2]{}
\providecommand{\surnamestart}{}
\providecommand{\surnameend}{}
\providecommand{\urlprefix}{Available at }
\providecommand{\url}[1]{\texttt{#1}}
\providecommand{\href}[2]{\texttt{#2}}
\providecommand{\urlalt}[2]{\href{#1}{#2}}
\providecommand{\doi}[1]{doi:\urlalt{http://dx.doi.org/#1}{#1}}
\providecommand{\bibinfo}[2]{#2}

\bibitemdeclare{article}{KeY2005}
\bibitem{KeY2005}
\bibinfo{author}{Wolfgang \surnamestart Ahrendt\surnameend},
  \bibinfo{author}{Thomas \surnamestart Baar\surnameend},
  \bibinfo{author}{Bernhard \surnamestart Beckert\surnameend},
  \bibinfo{author}{Richard \surnamestart Bubel\surnameend},
  \bibinfo{author}{Martin \surnamestart Giese\surnameend},
  \bibinfo{author}{Reiner \surnamestart H{\"a}hnle\surnameend},
  \bibinfo{author}{Wolfram \surnamestart Menzel\surnameend},
  \bibinfo{author}{Wojciech \surnamestart Mostowski\surnameend},
  \bibinfo{author}{Andreas \surnamestart Roth\surnameend},
  \bibinfo{author}{Steffen \surnamestart Schlager\surnameend} \&
  \bibinfo{author}{Peter~H. \surnamestart Schmitt\surnameend}
  (\bibinfo{year}{2005}): \emph{\bibinfo{title}{The {KeY} Tool}}.
\newblock {\sl \bibinfo{journal}{Software and System Modeling}}
  \bibinfo{volume}{4}(\bibinfo{number}{1}), pp. \bibinfo{pages}{32--54},
  \doi{10.1007/s10270-004-0058-x}.

\bibitemdeclare{book}{DBLP:series/txtcs/BertotC04}
\bibitem{DBLP:series/txtcs/BertotC04}
\bibinfo{author}{Yves \surnamestart Bertot\surnameend} \&
  \bibinfo{author}{Pierre \surnamestart Cast{\'{e}}ran\surnameend}
  (\bibinfo{year}{2004}): \emph{\bibinfo{title}{Interactive Theorem Proving and
  Program Development - Coq'Art: The Calculus of Inductive Constructions}}.
\newblock \bibinfo{series}{Texts in Theoretical Computer Science. An {EATCS}
  Series}, \bibinfo{publisher}{Springer}, \doi{10.1007/978-3-662-07964-5}.

\bibitemdeclare{inproceedings}{DBLP:conf/pldi/BohrerTMMP18}
\bibitem{DBLP:conf/pldi/BohrerTMMP18}
\bibinfo{author}{Brandon \surnamestart Bohrer\surnameend},
  \bibinfo{author}{Yong~Kiam \surnamestart Tan\surnameend},
  \bibinfo{author}{Stefan \surnamestart Mitsch\surnameend},
  \bibinfo{author}{Magnus~O. \surnamestart Myreen\surnameend} \&
  \bibinfo{author}{Andr{\'{e}} \surnamestart Platzer\surnameend}
  (\bibinfo{year}{2018}): \emph{\bibinfo{title}{{VeriPhy}: Verified Controller
  Executables from Verified Cyber-Physical System Models}}.
\newblock In \bibinfo{editor}{Dan \surnamestart Grossman\surnameend}, editor:
  {\sl \bibinfo{booktitle}{Proceedings of the 39th {ACM} {SIGPLAN} Conference
  on Programming Language Design and Implementation, {PLDI} 2018}},
  \bibinfo{publisher}{{ACM}}, pp. \bibinfo{pages}{617--630},
  \doi{10.1145/3192366.3192406}.

\bibitemdeclare{inproceedings}{DBLP:conf/itp/FultonMBP17}
\bibitem{DBLP:conf/itp/FultonMBP17}
\bibinfo{author}{Nathan \surnamestart Fulton\surnameend},
  \bibinfo{author}{Stefan \surnamestart Mitsch\surnameend},
  \bibinfo{author}{Brandon \surnamestart Bohrer\surnameend} \&
  \bibinfo{author}{Andr{\'{e}} \surnamestart Platzer\surnameend}
  (\bibinfo{year}{2017}): \emph{\bibinfo{title}{Bellerophon: Tactical Theorem
  Proving for Hybrid Systems}}.
\newblock In \bibinfo{editor}{Mauricio \surnamestart
  Ayala{-}Rinc{\'{o}}n\surnameend} \& \bibinfo{editor}{C{\'{e}}sar~A.
  \surnamestart Mu{\~{n}}oz\surnameend}, editors: {\sl
  \bibinfo{booktitle}{ITP}}, {\sl \bibinfo{series}{LNCS}}
  \bibinfo{volume}{10499}, \bibinfo{publisher}{Springer}, pp.
  \bibinfo{pages}{207--224}, \doi{10.1007/978-3-319-66107-0_14}.

\bibitemdeclare{inproceedings}{DBLP:conf/cade/FultonMQVP15}
\bibitem{DBLP:conf/cade/FultonMQVP15}
\bibinfo{author}{Nathan \surnamestart Fulton\surnameend},
  \bibinfo{author}{Stefan \surnamestart Mitsch\surnameend},
  \bibinfo{author}{Jan-David \surnamestart Quesel\surnameend},
  \bibinfo{author}{Marcus \surnamestart V{\"o}lp\surnameend} \&
  \bibinfo{author}{Andr{\'{e}} \surnamestart Platzer\surnameend}
  (\bibinfo{year}{2015}): \emph{\bibinfo{title}{{KeYmaera X}: An Axiomatic
  Tactical Theorem Prover for Hybrid Systems}}.
\newblock In \bibinfo{editor}{Amy \surnamestart Felty\surnameend} \&
  \bibinfo{editor}{Aart \surnamestart Middeldorp\surnameend}, editors: {\sl
  \bibinfo{booktitle}{CADE}}, {\sl \bibinfo{series}{LNCS}}
  \bibinfo{volume}{9195}, \bibinfo{publisher}{Springer},
  \bibinfo{address}{Berlin}, pp. \bibinfo{pages}{527--538},
  \doi{10.1007/978-3-319-21401-6_36}.

\bibitemdeclare{inproceedings}{DBLP:conf/icse/Leino04}
\bibitem{DBLP:conf/icse/Leino04}
\bibinfo{author}{K.~Rustan~M. \surnamestart Leino\surnameend}
  (\bibinfo{year}{2013}): \emph{\bibinfo{title}{Developing verified programs
  with {Dafny}}}.
\newblock In \bibinfo{editor}{David \surnamestart Notkin\surnameend},
  \bibinfo{editor}{Betty H.~C. \surnamestart Cheng\surnameend} \&
  \bibinfo{editor}{Klaus \surnamestart Pohl\surnameend}, editors: {\sl
  \bibinfo{booktitle}{35th Int. Conf. on Software Engineering, {ICSE} '13, San
  Francisco, CA, USA, May 18-26, 2013}}, \bibinfo{publisher}{{IEEE} Computer
  Soc.}, pp. \bibinfo{pages}{1488--1490}, \doi{10.1109/ICSE.2013.6606754}.

\bibitemdeclare{inproceedings}{DBLP:journals/corr/LeinoW14}
\bibitem{DBLP:journals/corr/LeinoW14}
\bibinfo{author}{K.~Rustan~M. \surnamestart Leino\surnameend} \&
  \bibinfo{author}{Valentin \surnamestart W{\"{u}}stholz\surnameend}
  (\bibinfo{year}{2014}): \emph{\bibinfo{title}{The {Dafny} Integrated
  Development Environment}}.
\newblock In \bibinfo{editor}{Catherine \surnamestart Dubois\surnameend},
  \bibinfo{editor}{Dimitra \surnamestart Giannakopoulou\surnameend} \&
  \bibinfo{editor}{Dominique \surnamestart M{\'{e}}ry\surnameend}, editors:
  {\sl \bibinfo{booktitle}{Proceedings 1st Workshop on Formal Integrated
  Development Environment, {F-IDE} 2014, Grenoble, France, April 6, 2014.}},
  {\sl \bibinfo{series}{{EPTCS}}} \bibinfo{volume}{149}, pp.
  \bibinfo{pages}{3--15}, \doi{10.4204/EPTCS.149.2}.

\bibitemdeclare{inproceedings}{DBLP:conf/emsoft/LiuZZ11}
\bibitem{DBLP:conf/emsoft/LiuZZ11}
\bibinfo{author}{Jiang \surnamestart Liu\surnameend}, \bibinfo{author}{Naijun
  \surnamestart Zhan\surnameend} \& \bibinfo{author}{Hengjun \surnamestart
  Zhao\surnameend} (\bibinfo{year}{2011}): \emph{\bibinfo{title}{Computing
  semi-algebraic invariants for polynomial dynamical systems}}.
\newblock In \bibinfo{editor}{Samarjit \surnamestart Chakraborty\surnameend},
  \bibinfo{editor}{Ahmed \surnamestart Jerraya\surnameend},
  \bibinfo{editor}{Sanjoy~K. \surnamestart Baruah\surnameend} \&
  \bibinfo{editor}{Sebastian \surnamestart Fischmeister\surnameend}, editors:
  {\sl \bibinfo{booktitle}{Proceedings of the 11th International Conference on
  Embedded Software, {EMSOFT} 2011, part of the Seventh Embedded Systems Week,
  ESWeek 2011, Taipei, Taiwan, October 9-14, 2011}},
  \bibinfo{publisher}{{ACM}}, pp. \bibinfo{pages}{97--106},
  \doi{10.1145/2038642.2038659}.

\bibitemdeclare{inproceedings}{DBLP:conf/fide/MitschP16}
\bibitem{DBLP:conf/fide/MitschP16}
\bibinfo{author}{Stefan \surnamestart Mitsch\surnameend} \&
  \bibinfo{author}{Andr{\'{e}} \surnamestart Platzer\surnameend}
  (\bibinfo{year}{2016}): \emph{\bibinfo{title}{The {KeYmaera X} proof {IDE}:
  Concepts on usability in hybrid systems theorem proving}}.
\newblock In \bibinfo{editor}{Catherine \surnamestart Dubois\surnameend},
  \bibinfo{editor}{Paolo \surnamestart Masci\surnameend} \&
  \bibinfo{editor}{Dominique \surnamestart M{\'{e}}ry\surnameend}, editors:
  {\sl \bibinfo{booktitle}{3rd Workshop on Formal Integrated Development
  Environment}}, {\sl \bibinfo{series}{EPTCS}} \bibinfo{volume}{240}, pp.
  \bibinfo{pages}{67--81}, \doi{10.4204/EPTCS.240.5}.

\bibitemdeclare{article}{DBLP:journals/fmsd/MitschP16}
\bibitem{DBLP:journals/fmsd/MitschP16}
\bibinfo{author}{Stefan \surnamestart Mitsch\surnameend} \&
  \bibinfo{author}{Andr{\'{e}} \surnamestart Platzer\surnameend}
  (\bibinfo{year}{2016}): \emph{\bibinfo{title}{{ModelPlex}: Verified Runtime
  Validation of Verified Cyber-Physical System Models}}.
\newblock {\sl \bibinfo{journal}{Form. Methods Syst. Des.}}
  \bibinfo{volume}{49}(\bibinfo{number}{1-2}), pp. \bibinfo{pages}{33--74},
  \doi{10.1007/s10703-016-0241-z}.
\newblock \bibinfo{note}{Special issue of selected papers from RV'14}.

\bibitemdeclare{incollection}{DBLP:series/lncs/MitschP20}
\bibitem{DBLP:series/lncs/MitschP20}
\bibinfo{author}{Stefan \surnamestart Mitsch\surnameend} \&
  \bibinfo{author}{Andr{\'{e}} \surnamestart Platzer\surnameend}
  (\bibinfo{year}{2020}): \emph{\bibinfo{title}{A Retrospective on Developing
  Hybrid Systems Provers in the {KeYmaera} Family - {A} Tale of Three
  Provers}}.
\newblock In \bibinfo{editor}{Wolfgang \surnamestart Ahrendt\surnameend},
  \bibinfo{editor}{Bernhard \surnamestart Beckert\surnameend},
  \bibinfo{editor}{Richard \surnamestart Bubel\surnameend},
  \bibinfo{editor}{Reiner \surnamestart H{\"{a}}hnle\surnameend} \&
  \bibinfo{editor}{Matthias \surnamestart Ulbrich\surnameend}, editors: {\sl
  \bibinfo{booktitle}{Deductive Software Verification: Future Perspectives -
  Reflections on the Occasion of 20 Years of {KeY}}}, {\sl
  \bibinfo{series}{LNCS}} \bibinfo{volume}{12345},
  \bibinfo{publisher}{Springer}, pp. \bibinfo{pages}{21--64},
  \doi{10.1007/978-3-030-64354-6_2}.

\bibitemdeclare{article}{DBLP:journals/sttt/MullerMRSP18}
\bibitem{DBLP:journals/sttt/MullerMRSP18}
\bibinfo{author}{Andreas \surnamestart M{\"{u}}ller\surnameend},
  \bibinfo{author}{Stefan \surnamestart Mitsch\surnameend},
  \bibinfo{author}{Werner \surnamestart Retschitzegger\surnameend},
  \bibinfo{author}{Wieland \surnamestart Schwinger\surnameend} \&
  \bibinfo{author}{Andr{\'{e}} \surnamestart Platzer\surnameend}
  (\bibinfo{year}{2018}): \emph{\bibinfo{title}{Tactical Contract Composition
  for Hybrid System Component Verification}}.
\newblock {\sl \bibinfo{journal}{STTT}}
  \bibinfo{volume}{20}(\bibinfo{number}{6}), pp. \bibinfo{pages}{615--643},
  \doi{10.1007/s10009-018-0502-9}.
\newblock \bibinfo{note}{Special issue for selected papers from FASE'17}.

\bibitemdeclare{inproceedings}{DBLP:conf/types/Nipkow02}
\bibitem{DBLP:conf/types/Nipkow02}
\bibinfo{author}{Tobias \surnamestart Nipkow\surnameend}
  (\bibinfo{year}{2002}): \emph{\bibinfo{title}{Structured Proofs in
  {Isar/HOL}}}.
\newblock In \bibinfo{editor}{Herman \surnamestart Geuvers\surnameend} \&
  \bibinfo{editor}{Freek \surnamestart Wiedijk\surnameend}, editors: {\sl
  \bibinfo{booktitle}{Types for Proofs and Programs, 2nd Int. Workshop, {TYPES}
  2002, Berg en Dal, The Netherlands, April 24-28, 2002, Selected Papers}},
  {\sl \bibinfo{series}{LNCS}} \bibinfo{volume}{2646},
  \bibinfo{publisher}{Springer}, pp. \bibinfo{pages}{259--278},
  \doi{10.1007/3-540-39185-1_15}.

\bibitemdeclare{book}{DBLP:books/sp/NipkowPW02}
\bibitem{DBLP:books/sp/NipkowPW02}
\bibinfo{author}{Tobias \surnamestart Nipkow\surnameend},
  \bibinfo{author}{Lawrence~C. \surnamestart Paulson\surnameend} \&
  \bibinfo{author}{Markus \surnamestart Wenzel\surnameend}
  (\bibinfo{year}{2002}): \emph{\bibinfo{title}{Isabelle/HOL - {A} Proof
  Assistant for Higher-Order Logic}}.
\newblock {\sl \bibinfo{series}{LNCS}} \bibinfo{volume}{2283},
  \bibinfo{publisher}{Springer}, \doi{10.1007/3-540-45949-9}.

\bibitemdeclare{inproceedings}{DBLP:conf/cade/Paulson10}
\bibitem{DBLP:conf/cade/Paulson10}
\bibinfo{author}{Lawrence~C. \surnamestart Paulson\surnameend}
  (\bibinfo{year}{2010}): \emph{\bibinfo{title}{Three Years of Experience with
  Sledgehammer, a Practical Link between Automatic and Interactive Theorem
  Provers}}.
\newblock In \bibinfo{editor}{Renate~A. \surnamestart Schmidt\surnameend},
  \bibinfo{editor}{Stephan \surnamestart Schulz\surnameend} \&
  \bibinfo{editor}{Boris \surnamestart Konev\surnameend}, editors: {\sl
  \bibinfo{booktitle}{Proceedings of the 2nd Workshop on Practical Aspects of
  Automated Reasoning, PAAR-2010, Edinburgh, Scotland, UK, July 14, 2010}},
  {\sl \bibinfo{series}{EPiC Series}}~\bibinfo{volume}{9},
  \bibinfo{publisher}{EasyChair}, pp. \bibinfo{pages}{1--10},
  \doi{10.29007/36dt}.

\bibitemdeclare{article}{DBLP:journals/jar/Platzer17}
\bibitem{DBLP:journals/jar/Platzer17}
\bibinfo{author}{Andr{\'{e}} \surnamestart Platzer\surnameend}
  (\bibinfo{year}{2017}): \emph{\bibinfo{title}{A Complete Uniform Substitution
  Calculus for Differential Dynamic Logic}}.
\newblock {\sl \bibinfo{journal}{J. Autom. Reas.}}
  \bibinfo{volume}{59}(\bibinfo{number}{2}), pp. \bibinfo{pages}{219--265},
  \doi{10.1007/s10817-016-9385-1}.

\bibitemdeclare{book}{Platzer18}
\bibitem{Platzer18}
\bibinfo{author}{Andr{\'{e}} \surnamestart Platzer\surnameend}
  (\bibinfo{year}{2018}): \emph{\bibinfo{title}{Logical Foundations of
  Cyber-Physical Systems}}.
\newblock \bibinfo{publisher}{Springer}, \bibinfo{address}{Cham},
  \doi{10.1007/978-3-319-63588-0}.

\bibitemdeclare{inproceedings}{DBLP:conf/cade/PlatzerQ08}
\bibitem{DBLP:conf/cade/PlatzerQ08}
\bibinfo{author}{Andr{\'{e}} \surnamestart Platzer\surnameend} \&
  \bibinfo{author}{Jan-David \surnamestart Quesel\surnameend}
  (\bibinfo{year}{2008}): \emph{\bibinfo{title}{{KeYmaera}: A Hybrid Theorem
  Prover for Hybrid Systems.}}
\newblock In \bibinfo{editor}{Alessandro \surnamestart Armando\surnameend},
  \bibinfo{editor}{Peter \surnamestart Baumgartner\surnameend} \&
  \bibinfo{editor}{Gilles \surnamestart Dowek\surnameend}, editors: {\sl
  \bibinfo{booktitle}{IJCAR}}, {\sl \bibinfo{series}{LNCS}}
  \bibinfo{volume}{5195}, \bibinfo{publisher}{Springer},
  \bibinfo{address}{Berlin}, pp. \bibinfo{pages}{171--178},
  \doi{10.1007/978-3-540-71070-7_15}.

\bibitemdeclare{article}{DBLP:journals/jacm/PlatzerT20}
\bibitem{DBLP:journals/jacm/PlatzerT20}
\bibinfo{author}{Andr{\'{e}} \surnamestart Platzer\surnameend} \&
  \bibinfo{author}{Yong~Kiam \surnamestart Tan\surnameend}
  (\bibinfo{year}{2020}): \emph{\bibinfo{title}{Differential Equation
  Invariance Axiomatization}}.
\newblock {\sl \bibinfo{journal}{J. ACM}}
  \bibinfo{volume}{67}(\bibinfo{number}{1}), pp. \bibinfo{pages}{6:1--6:66},
  \doi{10.1145/3380825}.

\bibitemdeclare{inproceedings}{DBLP:conf/icfem/RenshawLP11}
\bibitem{DBLP:conf/icfem/RenshawLP11}
\bibinfo{author}{David~W. \surnamestart Renshaw\surnameend},
  \bibinfo{author}{Sarah~M. \surnamestart Loos\surnameend} \&
  \bibinfo{author}{Andr{\'{e}} \surnamestart Platzer\surnameend}
  (\bibinfo{year}{2011}): \emph{\bibinfo{title}{Distributed Theorem Proving for
  Distributed Hybrid Systems}}.
\newblock In \bibinfo{editor}{Shengchao \surnamestart Qin\surnameend} \&
  \bibinfo{editor}{Zongyan \surnamestart Qiu\surnameend}, editors: {\sl
  \bibinfo{booktitle}{ICFEM}}, {\sl \bibinfo{series}{LNCS}}
  \bibinfo{volume}{6991}, \bibinfo{publisher}{Springer}, pp.
  \bibinfo{pages}{356--371}, \doi{10.1007/978-3-642-24559-6_25}.

\bibitemdeclare{article}{DBLP:journals/fmsd/SogokonMTCP}
\bibitem{DBLP:journals/fmsd/SogokonMTCP}
\bibinfo{author}{Andrew \surnamestart Sogokon\surnameend},
  \bibinfo{author}{Stefan \surnamestart Mitsch\surnameend},
  \bibinfo{author}{Yong~Kiam \surnamestart Tan\surnameend},
  \bibinfo{author}{Katherine \surnamestart Cordwell\surnameend} \&
  \bibinfo{author}{Andr{\'{e}} \surnamestart Platzer\surnameend}
  (\bibinfo{year}{2021}): \emph{\bibinfo{title}{Pegasus: Sound Continuous
  Invariant Generation}}.
\newblock {\sl \bibinfo{journal}{Form. Methods Syst. Des.}},
  \doi{10.1007/s10703-020-00355-z}.
\newblock \bibinfo{note}{Special issue for selected papers from FM'19}.

\bibitemdeclare{inproceedings}{DBLP:conf/tacas/TanP21}
\bibitem{DBLP:conf/tacas/TanP21}
\bibinfo{author}{Yong~Kiam \surnamestart Tan\surnameend} \&
  \bibinfo{author}{Andr{\'{e}} \surnamestart Platzer\surnameend}
  (\bibinfo{year}{2021}): \emph{\bibinfo{title}{Deductive Stability Proofs for
  Ordinary Differential Equations}}.
\newblock In \bibinfo{editor}{Jan~Friso \surnamestart Groote\surnameend} \&
  \bibinfo{editor}{Kim~Guldstrand \surnamestart Larsen\surnameend}, editors:
  {\sl \bibinfo{booktitle}{Tools and Algorithms for the Construction and
  Analysis of Systems - 27th International Conference, {TACAS} 2021, Held as
  Part of the European Joint Conferences on Theory and Practice of Software,
  {ETAPS} 2021, Luxembourg City, Luxembourg, March 27 - April 1, 2021,
  Proceedings, Part {II}}}, {\sl \bibinfo{series}{LNCS}}
  \bibinfo{volume}{12652}, \bibinfo{publisher}{Springer}, pp.
  \bibinfo{pages}{181--199}, \doi{10.1007/978-3-030-72013-1\_10}.

\bibitemdeclare{inproceedings}{DBLP:conf/tacas/TschannenFNP15}
\bibitem{DBLP:conf/tacas/TschannenFNP15}
\bibinfo{author}{Julian \surnamestart Tschannen\surnameend},
  \bibinfo{author}{Carlo~A. \surnamestart Furia\surnameend},
  \bibinfo{author}{Martin \surnamestart Nordio\surnameend} \&
  \bibinfo{author}{Nadia \surnamestart Polikarpova\surnameend}
  (\bibinfo{year}{2015}): \emph{\bibinfo{title}{AutoProof: Auto-Active
  Functional Verification of Object-Oriented Programs}}.
\newblock In \bibinfo{editor}{Christel \surnamestart Baier\surnameend} \&
  \bibinfo{editor}{Cesare \surnamestart Tinelli\surnameend}, editors: {\sl
  \bibinfo{booktitle}{Tools and Algorithms for the Construction and Analysis of
  Systems - 21st International Conference, {TACAS} 2015, London, UK, April
  11-18, 2015. Proceedings}}, {\sl \bibinfo{series}{LNCS}}
  \bibinfo{volume}{9035}, \bibinfo{publisher}{Springer}, pp.
  \bibinfo{pages}{566--580}, \doi{10.1007/978-3-662-46681-0}.

\bibitemdeclare{inproceedings}{DBLP:conf/icfem/WangZZ15}
\bibitem{DBLP:conf/icfem/WangZZ15}
\bibinfo{author}{Shuling \surnamestart Wang\surnameend},
  \bibinfo{author}{Naijun \surnamestart Zhan\surnameend} \&
  \bibinfo{author}{Liang \surnamestart Zou\surnameend} (\bibinfo{year}{2015}):
  \emph{\bibinfo{title}{An Improved {HHL} Prover: An Interactive Theorem Prover
  for Hybrid Systems}}.
\newblock In \bibinfo{editor}{Michael~J. \surnamestart Butler\surnameend},
  \bibinfo{editor}{Sylvain \surnamestart Conchon\surnameend} \&
  \bibinfo{editor}{Fatiha \surnamestart Za{\"{\i}}di\surnameend}, editors: {\sl
  \bibinfo{booktitle}{Formal Methods and Software Engineering - 17th
  International Conference on Formal Engineering Methods, {ICFEM} 2015, Paris,
  France, November 3-5, 2015, Proceedings}}, {\sl \bibinfo{series}{LNCS}}
  \bibinfo{volume}{9407}, \bibinfo{publisher}{Springer}, pp.
  \bibinfo{pages}{382--399}, \doi{10.1007/978-3-319-25423-4\_25}.

\bibitemdeclare{inproceedings}{DBLP:conf/aisc/Wenzel12}
\bibitem{DBLP:conf/aisc/Wenzel12}
\bibinfo{author}{Makarius \surnamestart Wenzel\surnameend}
  (\bibinfo{year}{2012}): \emph{\bibinfo{title}{{Isabelle}/{jEdit} - {A} Prover
  {IDE} within the {PIDE} Framework}}.
\newblock In \bibinfo{editor}{Johan \surnamestart Jeuring\surnameend},
  \bibinfo{editor}{John~A. \surnamestart Campbell\surnameend},
  \bibinfo{editor}{Jacques \surnamestart Carette\surnameend},
  \bibinfo{editor}{Gabriel~Dos \surnamestart Reis\surnameend},
  \bibinfo{editor}{Petr \surnamestart Sojka\surnameend},
  \bibinfo{editor}{Makarius \surnamestart Wenzel\surnameend} \&
  \bibinfo{editor}{Volker \surnamestart Sorge\surnameend}, editors: {\sl
  \bibinfo{booktitle}{Intelligent Computer Mathematics - 11th International
  Conference, {AISC} 2012, 19th Symp., Calculemus 2012, 5th Int. Workshop,
  {DML} 2012, 11th Int. Conf., {MKM} 2012, Systems and Projects, Held as Part
  of {CICM} 2012, Bremen, Germany, July 8-13, 2012. Proc.}}, {\sl
  \bibinfo{series}{LNCS}} \bibinfo{volume}{7362},
  \bibinfo{publisher}{Springer}, pp. \bibinfo{pages}{468--471},
  \doi{10.1007/978-3-642-31374-5}.

\bibitemdeclare{article}{DBLP:journals/jar/Wos87c}
\bibitem{DBLP:journals/jar/Wos87c}
\bibinfo{author}{Larry \surnamestart Wos\surnameend} (\bibinfo{year}{1987}):
  \emph{\bibinfo{title}{The Problem of Definition Expansion and Contraction}}.
\newblock {\sl \bibinfo{journal}{J. Autom. Reason.}}
  \bibinfo{volume}{3}(\bibinfo{number}{4}), pp. \bibinfo{pages}{433--435},
  \doi{10.1007/BF00247438}.

\end{thebibliography}
